\documentclass[rewiew,5p,times]{elsarticle}

\usepackage[pagewise]{lineno}
\usepackage{amsmath,linnum}

\usepackage{mathrsfs}
\usepackage{makecell}

\biboptions{sort&compress}

\usepackage{algorithm}
\usepackage{algorithmic}
\usepackage{amsmath}
\usepackage{CJKutf8}
\usepackage{graphicx}
\usepackage{float}
\usepackage{multirow}
\usepackage{amsfonts}
\usepackage{galois}
\usepackage{amssymb}
\usepackage{enumerate}
\usepackage{subfigure}
\newproof{proof}{Proof}

\usepackage{colortbl}
\usepackage{color, xcolor}
\usepackage{diagbox}
\usepackage{tcolorbox}
\usepackage{threeparttable}
\usepackage{bm}
\usepackage{hyperref}

\journal{Journal of Systems and Software}

\newcommand{\synote}[1]{\textcolor{black}{#1}}

\newcommand{\minorrevised}[1]{\textcolor{black}{#1}}

\begin{document}
\begin{CJK*}{UTF8}{gbsn}
	
\begin{frontmatter}

\title{A Comprehensive Empirical Investigation on Failure Clustering in Parallel Debugging}

\author[label1]{Yi Song}

\author[label1]{Xiaoyuan Xie\corref{cor1}}
\cortext[cor1]{Corresponding author.}
\ead{xxie@whu.edu.cn}

\author[label1]{Quanming Liu}

\author[label1]{Xihao Zhang}

\author[label1]{Xi Wu}

\address[label1]{School of Computer Science, Wuhan University, China\\}

\begin{abstract}
The clustering technique has attracted a lot of attention as a promising strategy for parallel debugging in multi-fault scenarios, this heuristic approach (i.e., failure indexing or fault isolation) enables developers to perform multiple debugging tasks simultaneously through dividing failed test cases into several disjoint groups. When using statement ranking representation to model failures for better clustering, several factors influence clustering effectiveness, including the risk evaluation formula (REF), the number of faults (NOF), the fault type (FT), and the number of successful test cases paired with one individual failed test case (NSP1F). In this paper, we present the first comprehensive empirical study of how these four factors influence clustering effectiveness. We conduct extensive controlled experiments on \synote{1060 faulty versions of 228 simulated faults and 141 real faults}, and the results reveal that: 1) GP19 is highly competitive across all REFs, 2) clustering effectiveness decreases as NOF increases, 3) higher clustering effectiveness is easier to achieve when a program contains only predicate faults, and 4) clustering effectiveness remains when the scale of NSP1F is reduced to 20\%.
\end{abstract}


\end{frontmatter}

\section{Introduction}
\label{sect1}

Programs often produce unexpected results that deviate from oracles during software testing, such anomalous behavior indicates that at least one fault resides in the program. However, locating these faults is generally labor-intensive and tedious in the debugging process~\cite{wong2016survey[1], xiaobo2018analysis[64]}. Generally, in multi-fault scenarios, there are two commonly adopted strategies:

\begin{itemize}
	
	\item \textbf{Sequential debugging.} Ignoring the linkage between failed test cases and faults, this strategy detects, localizes, and fixes one fault, and then reruns the test suite (TS, which contains all test cases) on the semi-repaired program under test (PUT) again, iterates these steps until a failure-free program is delivered.
	
	\item \textbf{Parallel debugging.} This strategy first mines the linkage that exists between failed test cases and faults, that is, divides all failed test cases into several disjoint fault-focused clusters through clustering techniques (with the goal of the failed test cases in a cluster to be triggered by the same root cause, and the failed test cases in different clusters to be triggered by different root causes), and combines each fault-focused cluster with all successful test cases to form several fault-focused TS, finally assigns them to different developers for parallel localization~\cite{jones2007debugging[2]}.
	
\end{itemize}

Many empirical studies have shown that sequential debugging does not perform well in localizing multiple faults~\cite{digiuseppe2011fault[3], digiuseppe2011influence[4], digiuseppe2015fault[5]}, while parallel debugging shows promise in this area. The core of parallel debugging lies in clustering. Only by properly capturing the linkage between failed test cases and faults, as well as heuristically dividing failed test cases, can a hunk of localization task be decomposed into several sub-tasks with high quality. However, most previous research in terms of parallel debugging concentrated on the localization process after clustering, with only a few studies investigating the clustering process, one of the most critical steps that may affect the overall parallel debugging performance. Several factors may affect the failure clustering step, but there is a lack of comprehensive empirical studies investigating these variables.

Therefore, in this paper, we conduct the first comprehensive empirical investigation, aiming at the clustering step by selecting four factors that could influence clustering effectiveness: the risk evaluation formula (REF) that represents failed test cases, the number of faults (NOF) and the fault type (FT) contained in the program, and the number of successful test cases paired with one individual failed test case (NSP1F), and further proposing four research questions as follows to guide our extensive experiments.

\begin{table}[!h]\footnotesize  %Table 3[h]
	\setlength{\belowcaptionskip}{3pt}
	\begin{center}
		\caption{\small{\synote{Abbreviations and their full forms}}}
		\label{tab:abbre}
		\resizebox{.85\columnwidth}{!}{
			\begin{tabular}{ll}
				\hline
				\textbf{Abbreviations} & \textbf{Full forms} \\   
				\hline
				REF & Risk Evaluation Formula \\
				NOF & the Number Of Faults \\
				FT & the Fault Type\\
				NSP1F & the Number of Successful test cases Paired with\\& ONE Failed test case \\
				TS & Test Suite \\
				PUT & Program Under Test \\
				AF & Assignment Fault \\
				PF & Predicate Fault \\
				\hline
		\end{tabular}}
	\end{center}
\end{table}

\vspace{3pt}
$\bullet$ \textbf{RQ1: Do different REFs have the same capability to representing failed test cases? }

Failed test cases are typically too unstructured and abstract to be used directly for clustering. Many approaches, such as coverage vector representation (CVR) and statement ranking representation (SRR), have been utilized to convert failed test cases into structured and mathematical forms. CVR is similar to T-proximity (Trace-proximity) in~\cite{liu2008systematic[6]}, in which a vector with a length equal to the number of executable statements in PUT is created to represent a failed test case, with the value of the $i^{\rm{th}}$ element being set to 1 if this failed test case covers the $i^{\rm{th}}$ statement, and 0 otherwise. SRR is similar to R-proximity (Rank-proximity) in~\cite{liu2008systematic[6]}, in which one failed test case and successful test cases are executed on PUT, and the coverage information of the program execution is collected and organized in the form of notations defined in spectrum-based fault localization (SBFL)~\cite{Xie2021[7]}. The coverage is then input into an REF to produce a ranking list that reflects statements' suspiciousness, which is employed to represent this failed test case finally. SRR has been proved to be superior to CVR in representing failed test cases~\cite{liu2008systematic[6]}, which has also been adopted by a number of previous research due to its advantage in translating a failed test case into a clustering-friendly proxy~\cite{2017Multiple[8], yu2015does[9], wang2014weighted[10], gao2017mseer[14]}.

In SRR, REF is used to produce a ranking list that contains the execution features of a failed test case. Obviously, a better REF should extract more discriminative features for failed test cases caused by different root causes, in other words, the distance between ranking lists that represent failed test cases triggered by different faults should be greater than the distance between ranking lists that represent failed test cases triggered by a same fault. However, almost all existing studies only simply chose a specific REF to generate the ranking list. To the best of our knowledge, no research has contrasted the capabilities of various REFs in representing failed test cases. To that end, we analyze 35 commonly-used REFs through extensive experiments in this RQ from this perspective.

\vspace{3pt}
$\bullet$ \textbf{RQ2: How NOF affects clustering effectiveness?}

Although it is difficult to know whether a faulty program contains a single fault or multiple faults exactly, we can intuitively infer the more faults it has, the more effort and time the debugging process will take~\cite{digiuseppe2011influence[4], xue2013significant[69]}. Many studies have investigated the effect of NOF on the effectiveness of fault localization techniques~\cite{digiuseppe2011influence[4], digiuseppe2015fault[5], jones2002visualization[11]}, but few have explored the influence of NOF on the clustering process. We analyze how clustering effectiveness changes as NOF grows in 2-bug, 3-bug, 4-bug, and 5-bug scenarios \synote{(i.e., programs that contain 2, 3, 4, and 5 bugs, respectively)}.

\vspace{3pt}
$\bullet$ \textbf{RQ3: Is clustering effectiveness affected by FT?}

In addition to NOF, FT is also an essential factor in the debugging process. Although the randomness and uncertainty of the programming process determine the diversity of the introduced faults, the most common FTs typically refer to assignment faults~\cite{jeffrey2008fault[12]} and predicate faults~\cite{xuan2016nopol[13]}. If a program has only assignment faults, only predicate faults, or both of them, how will the clustering effectiveness be affected? We discuss each of the three scenarios separately.

\vspace{3pt}
$\bullet$ \textbf{RQ4: Will clustering effectiveness be reduced using a lower NSP1F?}

When using SRR to represent failed test cases, almost all researchers pair one individual failed test case with all successful test cases~\cite{2017Multiple[8], yu2015does[9]} without giving any reason or explaining the rationality behind this strategy. If one failed test case is paired with \emph{part of} rather than \emph{all} successful test cases, the cost of debugging will be probably reduced, but will this reduction harm clustering effectiveness? We contrast the clustering effectiveness in five scenarios by pairing a failed test case with $X$ percent of successful test cases ($X$ = 100, 80, 60, 40, 20).

Furthermore, \synote{the distance metric, the estimation of the number of clusters and the assignment of initial medoids, as well as the clustering algorithm are also critical factors in determining clustering effectiveness in parallel debugging.} Gao and Wong have proposed a parallel debugging approach, MSeer~\cite{gao2017mseer[14]}, to solve the aforementioned concerns. In particular, \synote{they revised the traditional Kendall tau distance~\cite{kendall1990[80]}, presented an innovative strategy to assign initial medoids during predicting the number of clusters based on the mountain method~\cite{yager1994approximate[19], chiu1994fuzzy[20]}, and refined the K-medoids clustering algorithm~\cite{kaufman2009finding[45]}}. We will discuss our four research questions and conduct experiments based on MSeer due to its innovation and high effectiveness. \synote{A further introduction regarding MSeer is given in Section \ref{subsect2.3}.}

We create \synote{1060} faulty versions of \synote{nine} programs, $flex$, $grep$, $gzip$, $sed$, \synote{$Chart$, $Closure$, $Lang$, $Math$, and $Time$,} as our benchmark. The experimental results show that\footnote{\synote{The replication package of this empirical study is available at \href{https://github.com/yisongy/failureClustering}{this website}.}}:

\hangafter=1
\setlength{\hangindent}{2.7em}
1) GP19 \minorrevised{(the $19^{\rm{th}}$ formula evolved by Genetic Programming in~\cite{yoo2012evolving[41]})} is highly competitive across all REFs when representing failed test cases.

2) Clustering effectiveness decreases as NOF grows.

\hangafter=1
\setlength{\hangindent}{2.7em}
3) Higher clustering effectiveness is easier to achieve when a faulty program contains only predicate faults.

\hangafter=1
\setlength{\hangindent}{2.7em}
4) Clustering effectiveness remains when NSP1F is reduced to 20\%.

\vspace{3pt}
\textbf{The main contributions of this paper are as follows:}

\hangafter=1
\setlength{\hangindent}{2.7em}
1) Unlike previous studies that contrasted REFs from the perspective of fault localization effectiveness, we contrast 35 REFs (including the latest Crosstab, Dstar, and GP02, GP03, GP19 evolved by genetic programming) in terms of how well they represent failed test cases. We recommend GP19, an REF with strong competitiveness in extracting failed test cases' execution features for future researchers.

\hangafter=1
\setlength{\hangindent}{2.7em}
2) Our controlled experiments reveal that the effectiveness of clustering failed test cases will reduce when NOF increases.

\hangafter=1
\setlength{\hangindent}{2.7em}
3) We analyze two typical types of faults, assignment faults and predicate faults, and discover that it is easier to achieve higher clustering effectiveness when a program contains only predicate faults.

\hangafter=1
\setlength{\hangindent}{2.7em}
4) We pair 100\%, 80\%, 60\%, 40\%, and 20\% of successful test cases with one failed test case, and contrast the clustering effectiveness in these five scenarios. The findings indicate that cutting the scale of successful test cases has little effect on clustering effectiveness, suggesting a way worth trying to lower the cost of SRR representation for future researchers.

The remainder of this paper is organized as follows: Section~\ref{sect2} introduces the background knowledge. Section~\ref{sect3} describes the experimental dataset and setup. Section~\ref{sect4} analyzes the experimental results. Section~\ref{sect5} discusses some interesting topics. Section~\ref{sect6} is the threats to validity. Section~\ref{sect7} reports related works. Conclusions and directions for future work are proposed in Section~\ref{sect8}.

\section{Background}
\label{sect2}

We explain why clustering failed test cases is essential and present the rationale of parallel debugging in Section \ref{subsect2.1}. The principles and technical details of SRR are given in Section \ref{subsect2.2}, followed by a motivating example showing the application of SRR-based failure clustering in Section \ref{subsect2.3}.

\subsection{Why Clustering?}
\label{subsect2.1}

In general, the possibility of a program being faulty and the number of faults it contains are proportional to its size~\cite{wang2008[56]}. With the increasing volume and the explosive growth of code in modern software systems, most faulty programs usually have multiple faults.

In multi-fault scenarios, various failed test cases\footnote{In this paper, we use ``failed test case", ``anomalous execution", and ``failure" interchangeably.} may be caused by different faults. If failed test cases with distinct root causes are not divided properly, fault localization techniques could be confused by the impure test suite significantly, for example, SBFL techniques extract execution features of all faults guided by the impure spectrum information, which will lower the rank of each fault in the generated ranking list. According to Wang et al., \emph{failed test cases that are not related to specific fault are the main reason to reduce the effectiveness of SBFL}~\cite{wang2020[28]}, and similarly, Keller et al. have drawn a similar conclusion, \emph{when using SBFL techniques, the number of lines that need to be inspected can be reduced by high quality test cases that execute the bug}~\cite{keller2017critical[77]}. Therefore, the purpose of dividing failed test cases in a multi-fault scenario is to allow failed test cases with different root causes to target their corresponding faults separately, to put it another way, reduce the interferences among multiple faults in a program, enhance the pertinence of fault localization techniques and thus achieve parallel debugging.

Many researchers have attempted to employ the clustering technique to divide failed test cases~\cite{jones2007debugging[2], gao2017mseer[14], wu2020fatoc[23], golagha2019failure[70], digiuseppe2012concept[71]}. Ideally, failures caused by the same fault should be grouped into a cluster, then the failed test cases in a cluster are combined with all successful test cases to form a fault-focused TS targeting a specific fault, as defined in Formula \ref{equ1} and Formula \ref{equ2}. This strategy is often called failure indexing or fault isolation.

\begin{equation}
	\label{equ1}
	F_{t}=F_{1} \cup F_{2} \cup \cdots \cup F_{r}
\end{equation}

\begin{equation}
	\label{equ2}
	fault{\raisebox{0mm}{-}}focused\ TS_{i}=F_{i} \cup S (i = 1, 2, ..., r)
\end{equation}

Where $F_{t}$ and $S$ represent all failed test cases and all successful test cases in TS, respectively. $F_{1}$, $F_{2}$, ..., $F_{r}$ are generated fault-focused clusters, and $r$ is the number of clusters (which is expected to be equal to the number of faults).

Clustering failed test cases is a heuristic strategy for improving the pertinence of TS and the effectiveness of fault localization, this widely acknowledged method has been adopted by many previous studies in the field of multi-fault localization~\cite{gao2017mseer[14], podgurski2003automated[15], steimann2012improving[32]}.

It is vital to encode failed test cases in an intermediate representation due to their unfriendly form for clustering. Currently, the most widely used representation methods are aforementioned CVR and SRR. The technical details of SRR, which is employed to conduct experiments in this paper, are described below.

\subsection{Statement Ranking Representation}
\label{subsect2.2}

After TS have been executed on PUT, the coverage information of each test case that contains two components will be collected in SRR:

\begin{itemize}
	\item \textbf{Execution Path}: A binary vector that records which program entities (statements\footnote{We implement the statement granularity in our experiments, hence ``entity" and ``statement" are interchangeable hereafter.}, branches, functions, or basic blocks)~\cite{reps1997use[29], harrold2000empirical[30]} have been covered by a test case.  
	
	\item \textbf{Execution result}: A binary value denotes whether or not the actual output of a test case matches its expected output.
\end{itemize}

Suppose there is a PUT containing $j$ executable statements $s_{i}$ ($i$ = 1, 2, …, $j$) and a TS containing $p$ test cases $t_{i}$ ($i$ = 1, 2, …, $p$), the coverage generated by running TS against PUT should be a matrix of size $j$ × $p$. In SRR, the coverage gathered against a failed test case and successful test cases will be converted into spectrum information according to the notations defined in SBFL~\cite{Xie2021[7]}, as shown in Table \ref{tab:spectrum} \footnote{Also referred to as $a_{ef}$, $a_{nf}$, $a_{ep}$, $a_{np}$, $a_{e}$, $a_{n}$, $a_{p}$, $a_{f}$, $a$, respectively.}.

\begin{table}[!h]\footnotesize  %Table 3[h]
	\setlength{\belowcaptionskip}{3pt}
	\begin{center}
		\caption{\small{Notations in spectrum information}}
		\label{tab:spectrum}
		\begin{tabular}{ll}
			\hline
				\textbf{Notation} & \textbf{Meaning} \\   
			\hline
				$N_{CF}$ & the number of failed test cases covering a statement \\
				$N_{UF}$ & the number of failed test cases not covering a statement \\
				$N_{CS}$ & the number of successful test cases covering a statement \\
				$N_{US}$ & the number of successful test cases not covering a statement \\
				$N_{C}$ & the number of test cases covering a statement \\
				$N_{U}$ & the number of test cases not covering a statement \\
				$N_{S}$ & total number of successful test cases \\ 
				$N_{F}$ & total number of failed test cases \\ 
				$N$ & total number of test cases \\
			\hline
		\end{tabular}
	\end{center}
\end{table}

To incorporate several notations in spectrum information into a suspiciousness value that measures the risk of a statement being faulty, researchers have constructed a series of risk evaluation formulas. For example, Ochiai proposed by Abreu et al. is defined in Formula \ref{equ3}~\cite{abreu2006evaluation[31]}:

\begin{equation}
	\label{equ3}
	\text { $suspiciousness$ }_{\text {Ochiai }}=\frac{N_{C F}}{\sqrt{N_{F} N_{C}}}
\end{equation}

The statements \footnote{Unless otherwise specified, ``statement" refers to ``executable statement" in this paper.} in PUT are ranked according to their suspiciousness in descending order to deliver a ranking list. This type of ranking list, which is produced by an REF from spectrum information that reflects the execution features of a failed test case and successful test cases, is employed to represent this failed test case in SRR.

\subsection{\synote{Motivating Example}}
\label{subsect2.3}

\begin{figure}  
	\footnotesize 
	\graphicspath{}
	\centering
	\includegraphics[width = 8.5cm, height = 5.7cm]{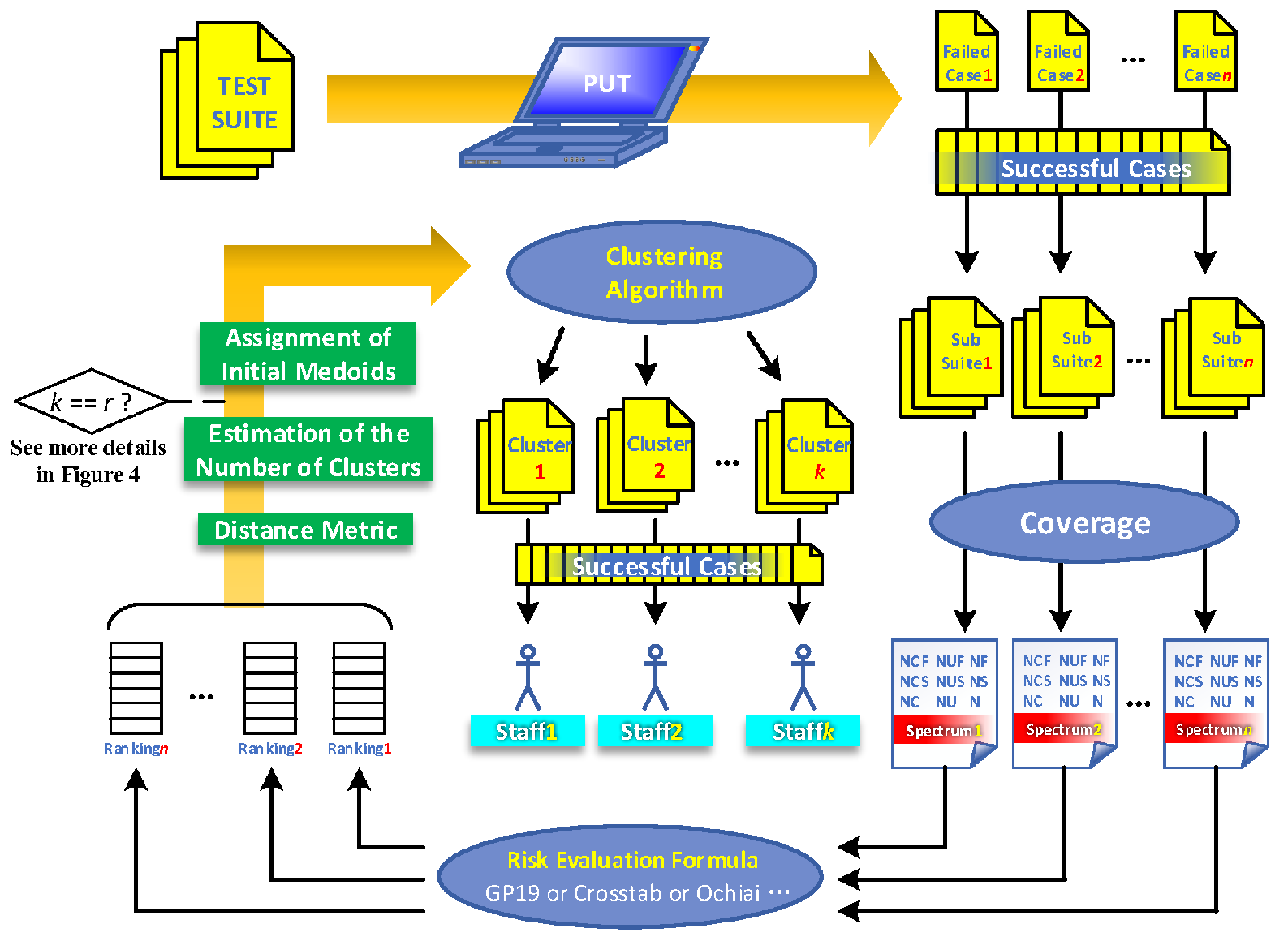}
	\caption{\synote{The workflow of SRR-based failure clustering}}
	\label{fig:workflow} 
\end{figure}

The workflow of SRR-based failure clustering is illustrated in Figure \ref{fig:workflow}. \synote{Test cases in the test suite can be determined as failed or successful after being executed against the program, according to the inconsistency or consistency between actual and expected outputs, respectively. Each of failed test cases will be combined with successful test cases and then be fed into a risk evaluation formula, for delivering a ranking list that could represent it in a mathematical form. Once fault-focused clusters are produced by clustering these ranking lists, they will be immediately sent to different handlers for the following step. It should be noted that after failed test cases have been transformed to ranking lists, it is necessary to preprocess such data by measuring distances between them, estimating the number of clusters, and assigning the initial medoids, and only after all of these procedures have been fulfilled can the clustering algorithm begin to work.} MSeer, an advanced framework for localizing multiple faults in parallel that alleviated these challenging jobs, has been proposed by Gao and Wong~\cite{gao2017mseer[14]}. \synote{Specifically, they 1) claimed that in the classic Kendall tau distance metric, discordant pairs of more suspicious statements should contribute more to the distance between two ranking lists, and proposed a modified distance metric based on this intuition; 2) assigned a potential value to each of failed test cases (ranking lists) based on data winsorization, and developed an algorithm to judge whether a failed test case should be chosen as one of medoids; 3) relieved the shortcoming of examining all possible combinations of data points as initial medoids that exists in the traditional K-medoids clustering algorithm.} We conduct our experiments based on MSeer because it has been recognized as one of the state-of-the-art parallel debugging techniques, along with its availability and reliability.

Let us use a motivating example to illustrate the details of SRR as well as demonstrate the promise of failure clustering. As shown in Table \ref{tab:example}, the PUT that contains 11 statements, is designed to calculate the product of the smaller two of the three numbers, in which two faults have been induced by statements $s_{6}$ and $s_{9}$, respectively. Give a TS containing 10 test cases: $t_{1}$ = \{1,2,4\},\ $t_{2}$ = \{4,3,2\},\ $t_{3}$ = \{3,2,4\},\ $t_{4}$ = \{5,1,6\},\ $t_{5}$ = \{2,6,5\},\ $t_{6}$ = \{6,5,1\},\ $t_{7}$ = \{7,5,8\},\ $t_{8}$ = \{5,7,3\},\ $t_{9}$ = \{8,1,2\},\ $t_{10}$ = \{8,6,9\}, six of them are labelled as $failed$ due to the unexpected outputs ($t_{3}$, $t_{4}$, $t_{5}$, $t_{7}$, $t_{8}$, $t_{10}$). The 11×10 matrix composed of rows $s_{1}$  to $s_{11}$  and columns $t_{1}$ to $t_{10}$ in Table \ref{tab:example} is the coverage obtained by running TS against PUT, where $t_{1}$ $\sim$ $t_{10}$ columns represent the execution paths of 10 test cases. The symbol ``·'' denotes that a test case covers an innocent statement, while ``$\blacktriangle$'' and ``$\triangle$'' denote that a test case covers the statements containing $Fault_{1}$ and $Fault_{2}$, respectively. The coverage information is reorganized to spectrum information according to the notations defined in Table \ref{tab:spectrum}, as shown in the 11×9 matrix composed of rows $s_{1}$  to $s_{11}$  and columns $N_{CF}$ to $N$ in Table \ref{tab:example}.

\begin{table*}[]\footnotesize
	\centering
	\setlength{\belowcaptionskip}{3pt}
	\caption{\label{tab:example} The sample PUT and its coverage against the given TS} 
	\begin{tabular}{|c|l|p{1pt}p{1pt}p{1pt}p{1pt}p{1pt}p{1pt}p{1pt}p{1pt}p{1pt}p{5pt}|p{3pt}p{3pt}p{3pt}p{3pt}p{3pt}p{3pt}p{3pt}p{3pt}p{5pt}|ccc|}
		\cline{1-24}
	    \multirow{2}{*}{\textbf{S}} & \multirow{2}{*}{\textbf{Program}} & \multicolumn{10}{c|}{\textbf{Coverage Information}} & \multicolumn{9}{c|}{\textbf{Spectrum Information}} &  \multicolumn{3}{c|}{\textbf{Suspiciousness}}\\
		\cline{3-24} & & $t_{1}$ & $t_{2}$ & \textbf{$t_{3}$} & \textbf{$t_{4}$} & \textbf{$t_{5}$} & $t_{6}$& \textbf{$t_{7}$} & \textbf{$t_{8}$} & $t_{9}$ & \textbf{$t_{10}$}  &  $N_{CF}$ & $N_{UF}$ & $N_{CS}$ & $N_{US}$ & $N_{C}$ & $N_{U}$ & $N_{S}$ & $N_{F}$ & $N$ & \tiny{$F_{t} \cup S$} & \tiny{$F_1 \cup S$} & \tiny{$F_2 \cup S$}\\
		\cline{1-24}
		\textbf{$s_{1}$}  & input a, b, c                                                & ·                    & ·                    & ·                    & ·                    & ·                    & ·                    & ·                    & ·                    & ·                    & ·   & 6       & 0           & 4           & 0           & 10         & 0          & 4         & 6         & 10   &   0.77    &   0.58 &   0.71           \\
		\textbf{$s_{2}$}  & if (a \textless \ b):                                           & ·                    & ·                    & ·                    & ·                    & ·                    & ·                    & ·                    & ·                    & ·                    & ·  & 6       & 0           & 4            & 0           & 10         & 0         & 4         & 6         & 10    & 0.77  &     0.58 &   0.71             \\
		\textbf{$s_{3}$}  & \quad if (b \textless \ c):                                           & ·                    &                      &                      &                      & ·                    &                      &                      & ·                    &                      &     & 2       & 4          & 1          & 3           & 3          & 7          & 4          & 6         & 10 & 0.47   &   0.82 &   0              \\
		\textbf{$s_{4}$} & \qquad z = a * b                                                    & ·                    &                      &                      &                      &                      &                      &                      &                      &                      &    & 0      & 6           & 1           & 3           & 1          & 9         & 4          & 6         & 10   & 0    &   0 &   0            \\
		\textbf{$s_{5}$} & \quad else:                                                         &                      &                      &                      &                      & ·                    &                      &                      & ·                    &                      &     & 2       & 4          & 0          & 4         & 2       & 8       & 4     & 6         & 10   & 0.58 &   \textbf{1} &   0             \\
		\rowcolor{gray!30}\textbf{$s_{6}$}  & \qquad z = b * c  ~\scriptsize{//$Fault_1$  ~\checkmark z = a * c}                                &                      &                      &                      &                      & $\blacktriangle$                    &                      &                      & $\blacktriangle$                    &                      &      & 2       & 4         & 0         & 4        & 2        & 8       & 4      & 6       & 10    & 0.58  &   \textbf{1} &   0        \\
		\textbf{$s_{7}$} & else:                                                         &                      & ·                    & ·                    & ·                    &                      & ·                    & ·                    &                      & ·                    & ·  & 4       & 2          & 3         & 1         & 7      & 3         & 4         & 6         & 10      & 0.62   &   0 &   0.76           \\
		\textbf{$s_{8}$} & \quad if  (a \textless c)                                          &                      & ·                    & ·                    & ·                    &                      & ·                    & ·                    &                      & ·                    & ·     &   4     & 2        & 3       & 1        & 7         & 3        & 4       & 6       & 10   & 0.62    &   0 &   0.76     \\
		\rowcolor{gray!30}\textbf{$s_{9}$}  & \qquad z = a * c ~\scriptsize{//$Fault_2$  ~\checkmark z = a * b}                                &                      &                      & $\triangle$                    & $\triangle$                    &                      &                      & $\triangle$                    &                      &                      & $\triangle$      & 4      & 2         & 0       & 4       & 4         & 6        & 4       & 6         & 10   & \textbf{0.82}   &   0 &  \textbf{ 1}       \\
		\textbf{$s_{10}$} & \quad else                                                         &                      & ·                    &                      &                      &                      & ·                    &                      &                      & ·                    &       & 0      & 6          & 3        & 1          & 3          & 7        & 4      & 6       & 10   & 0     &   0 &   0        \\
		\textbf{$s_{11}$} & \qquad z = b * c                                                    &                      & ·                    &                      &                      &                      & ·                    &                      &                      & ·                    &        & 0        & 6           & 3          & 1          & 3         & 7          & 4          & 6        & 10   & 0    &   0 &   0     \\  \cline{1-24}
	\end{tabular}
\end{table*}

Each statement's suspiciousness is then generated by Ochiai, as shown in column $F_{t} \cup S$ in Table \ref{tab:example}. We can immediately sort these statements in descending order of suspiciousness, and then get a ranking list of them: {$s_{9}$, $s_{1}$, $s_{2}$, $s_{7}$, $s_{8}$, $s_{5}$, $s_{6}$, $s_{3}$, $s_{4}$, $s_{10}$, $s_{11}$}. The statement $s_{9}$ containing $Fault_{2}$ has the highest suspiciousness of 0.82, hence it will be inspected first. However, the statement $s_{6}$ containing $Fault_{1}$ is ranked seventh, innocent statements $s_{1}$, $s_{2}$, $s_{7}$, $s_{8}$ and $s_{5}$ will be examined before $s_{6}$. This simple example reveals that the impure TS has a limited capability to delivering a promising fault localization output.

\synote{Now we depict how fault localization effectiveness will be improved by grouping failed test cases into distinct fault-focused clusters. This is also a step-by-step elaboration of Figure \ref{fig:workflow}.}

\begin{itemize}

\item \synote{\textbf{For the failure representation.} We employ SRR to represent all six failed test cases. Take $t_{5}$ as an example. Pairing $t_{5}$ with $S$ to form a failure-specific TS, $t_{5} \cup S$, executing this TS on PUT to obtain coverage and convert it into spectrum information\footnote{This failure-specific TS's coverage and the corresponding spectrum information are omitted due to limited space.}, and then utilizing a risk evaluation formula (e.g., Ochiai) to incorporate the spectrum information for obtaining each statement's suspiciousness, finally, a ranking list can be produced to represent $t_{5}$, as shown in Table \ref{tab:t5andS}, which will be invoked in the subsequent clustering process as a proxy of $t_{5}$.}
\begin{table}[]\footnotesize
	\centering
	\setlength{\belowcaptionskip}{3pt}
	\caption{\label{tab:t5andS} Statements’ suspiciousness calculated by Ochiai in the sample PUT against $t_{5} \cup S$ and the corresponding ranking list} 
	\begin{tabular}{p{1.6cm}p{0.19cm}p{0.19cm}p{0.19cm}p{0.19cm}p{0.19cm}p{0.19cm}p{0.19cm}p{0.19cm}p{0.19cm}p{0.19cm}p{0.19cm}}
		\hline
		\textbf{Statement}               & \bm{$s_{1}$} & \bm{$s_{2}$} & \bm{$s_{3}$} & \bm{$s_{4}$} & \bm{$s_{5}$} & \bm{$s_{6}$} & \bm{$s_{7}$} & \bm{$s_{8}$} & \bm{$s_{9}$} & \bm{$s_{10}$} & \bm{$s_{11}$} \\ \hline
		Suspiciousness & 0.45        & 0.45        & 0.71        & 0           & 1       & 1        & 0       & 0        & 0         & 0         & 0         \\
		Ranking list                & 4                    & 4                    & 3                    & 6                    & 1                    & 1                    & 6                    & 6                    & 6                    & 6                     & 6         \\   \hline          
	\end{tabular}
\end{table}
\synote{It should be noted that there are many ways for producing a ranking list according to statements’ suspiciousness~\cite{huang2013empirical[26]}. Considering the intuition that a ranking list should clearly reflect the priority of a statement being inspected, as well as other previous studies’ experience~\cite{huang2013empirical[26]}, we adopt the following ranking strategy: if several statements with the same suspiciousness form a $Tie$~\cite{xu2011ties[46]}, the rankings of all statements in the $Tie$ will be set to the beginning position of this $Tie$.}

\item \synote{\textbf{For the distance metric.} Given two ranking lists that represent failed test cases, the classical Kendall tau distance counts the number of pairwise disagreements between them. Considering the characteristic of ranking lists in the context of failure representation, discordant pairs of more risky statements (i.e., at lower positions in the ranking lists) should be paid more attention. Based on this intuition, we use the revised Kendall tau distance, which takes the reciprocal of the position of statements in the discordant pairs~\cite{gao2017mseer[14]}, to measure the similarity between each pair of failed test cases.}

\item \synote{\textbf{For the estimation of the number of clusters and the assignment of initial medoids.} We assign a potential value for each failed test case according to the density of its surrounding, to reflect the possibility of it being set as a medoid, and the failed test case with the highest potential value will be selected as the first medoid. Then, all failed test cases' potential values will be updated based on how far they are from the newest medoid. Repeating these steps iteratively until the highest potential value falls within a predefined threshold, and as a consequence of which, the number of clusters and initial medoids can be determined at the same time~\cite{gao2017mseer[14]}.}

\item \synote{\textbf{For the clustering algorithm.} The K-medoids clustering approach sets practical (not virtual) data points as medoids, aiming at minimizing the distance between failed test cases and the medoid of the cluster where they reside. Its traditional version suffers from two tricky problems, namely, the difficulty of choosing a proper distance metric and the overhead caused by examining all possible combinations of data samples as initial medoids. The aforementioned two strategies can properly handle these two points, respectively, thus an improved K-medoids algorithm can be delivered and used in our failure clustering~\cite{gao2017mseer[14]}. In the motivating example, failed test cases $t_{5}$ and $t_{8}$ are triggered by $Fault_{1}$, and $t_{3}$, $t_{4}$, $t_{7}$, and $t_{10}$ are triggered by $Fault_{2}$. Ideally, the clustering results should be $F_{1}$ = \{$t_{5}$, $t_{8}$\}, $F_{2}$ = \{$t_{3}$, $t_{4}$, $t_{7}$, $t_{10}$\}\footnote{\synote{For more details about the distance metric, the estimation of the number of clusters and the assignment of initial medoids, and the clustering algorithm, please refer to \cite{gao2017mseer[14]}.}}.}

\item \synote{\textbf{For the bug triage.} Two fault-focused TSs, $F_{1} \cup S$, $F_{2} \cup S$, can be produced by combining $F_{1}$ and $F_{2}$ with all successful test cases $S$ separately, and two sets of spectrum information can be collected by executing them on PUT accordingly\footnote{These two fault-focused TSs' coverage and the corresponding spectrum information are omitted due to limited space.}. The suspiciousness of statements calculated by Ochiai using these two sets of spectrum information is shown in columns $F_{1} \cup S$ and $F_{2} \cup S$ in Table \ref{tab:example}, respectively. In the ranking list produced against $F_{1} \cup S$, the statement $s_{6}$ where $Fault_{1}$ lies in is given the highest suspiciousness, while in the ranking list produced against $F_{2} \cup S$, the statement $s_{9}$ where $Fault_{2}$ lies in is given the highest suspiciousness. Surprisingly, each faulty statement appears at the top of the corresponding ranking list. Guided by such fault localization outputs with strong pertinence, a developer (in sequential debugging), or two developers (in parallel debugging), only need(s) to inspect at most three statements (the suspiciousness of $s_{5}$ and $s_{6}$ calculated against $F_{1} \cup S$ is identical) for localizing all two faults. However, at least six statements have to be examined for finding two faults in the confusing ranking list produced without clustering failed test cases.}

\end{itemize}

This motivating example not only highlights the promise of clustering failed test cases but also indicates some key factors in such a process:  the risk evaluation formula (\textbf{\emph{REF}}) that produces ranking lists to representing failed test cases, the number of successful test cases paired with one individual failed test case (\textbf{\emph{NSP1F}}), may influence clustering effectiveness. Furthermore, considering that the effect of the number of faults (\textbf{\emph{NOF}}) and the fault type (\textbf{\emph{FT}}) in PUT on software debugging has caught the attention of fault localization communities~\cite{digiuseppe2011influence[4], digiuseppe2015fault[5], jones2002visualization[11]}, we conjecture these two points are also likely to affect the results of clustering. We conduct extensive controlled experiments to explore how these four factors affect the clustering process in the next section.

\section{EXPERIMENTAL SETUP}
\label{sect3}

Section \ref{subsect3.1} provides the dataset used in our experiments and the mechanism for generating multi-fault versions via mutation-based strategies. Section \ref{subsect3.2} describes experimental setups for four RQs. Section \ref{subsect3.3} introduces four metrics for evaluating the experimental results.

\subsection{The generation of faulty versions}
\label{subsect3.1}

\iffalse
We choose four benchmark programs from SIR~\cite{SIR[33]}: $flex$ (lexical analyzer), $grep$ (file patterns searcher), $gzip$ (data compressor), and $sed$ (text processor), as shown in Table \ref{tab:benchmark}.
\fi

We choose four benchmark programs from SIR~\cite{SIR[33]}: $flex$, $grep$, $gzip$, and $sed$, \synote{and five benchmark programs from Defects4J~\cite{just2014defects4j[81]}: $Chart$, $Closure$, $Lang$, $Math$, and $Time$}, \synote{for the generation of multi-fault versions,} as shown in Table \ref{tab:benchmark}.

\iffalse
\begin{table}[]\footnotesize
	\centering
	\setlength{\belowcaptionskip}{3pt}
	\caption{\label{tab:benchmark} Subject Programs} 
	\begin{tabular}{ccllcccl}
		\hline
		\textbf{Project} & \multicolumn{3}{c}{\textbf{Version}} & \textbf{LOC} & \textbf{No. of test cases} & \multicolumn{2}{c}{\textbf{\begin{tabular}[c]{@{}c@{}}No. of faults\\ (AF / PF)\end{tabular}}}  \\ \hline
		flex             & \multicolumn{3}{c}{2.5.3}            & 14477        & 525                        & \multicolumn{2}{c}{30 / 46}  \\
		grep             & \multicolumn{3}{c}{2.4}              & 13463        & 470                        & \multicolumn{2}{c}{27 / 20}  \\
		gzip             & \multicolumn{3}{c}{1.2.2}            & 7343         & 213                        & \multicolumn{2}{c}{24 / 20}   \\
		sed              & \multicolumn{3}{c}{3.02}             & 10225        & 363                        & \multicolumn{2}{c}{21 / 40}  \\ \hline	\end{tabular}
\end{table}
\fi

\begin{table}[]\footnotesize
	\centering
	\setlength{\belowcaptionskip}{3pt}
	\caption{\label{tab:benchmark} \synote{Subject Programs}} 
	\begin{tabular}{lcccl}
		\hline
		\textbf{Project} & \textbf{Version} & \textbf{kLOC} & \textbf{No. of faults} & \textbf{Description}  \\ \hline
		flex             & 2.5.3            & 14.5        & 30AF + 46PF   & lexical analyzer   \\
		grep             & 2.4              & 13.5        & 27AF + 20PF  & file patterns searcher  \\
		gzip             & 1.2.2            & 7.3         & 24AF + 20PF       & data compressor   \\
		sed              & 3.02             & 10.2        & 21AF + 40PF   & text processor  \\
		Chart & 2.0.0 & 96.3 & 18 & Chart library \\
		Closure & 2.0.0 & 90.2 & 36 & Closure compiler \\
		Lang & 2.0.0 & 22.1 & 38 & Apache commons-lang \\
		Math & 2.0.0 & 85.5 & 29 & Apache commons-math \\
		Time & 2.0.0 & 28.4 & 20 & Date and time library \\
		 \hline	\end{tabular}
\end{table}

\begin{figure}
	\footnotesize 
	\graphicspath{}  
	\centering  
	\includegraphics[width=0.33\textwidth]{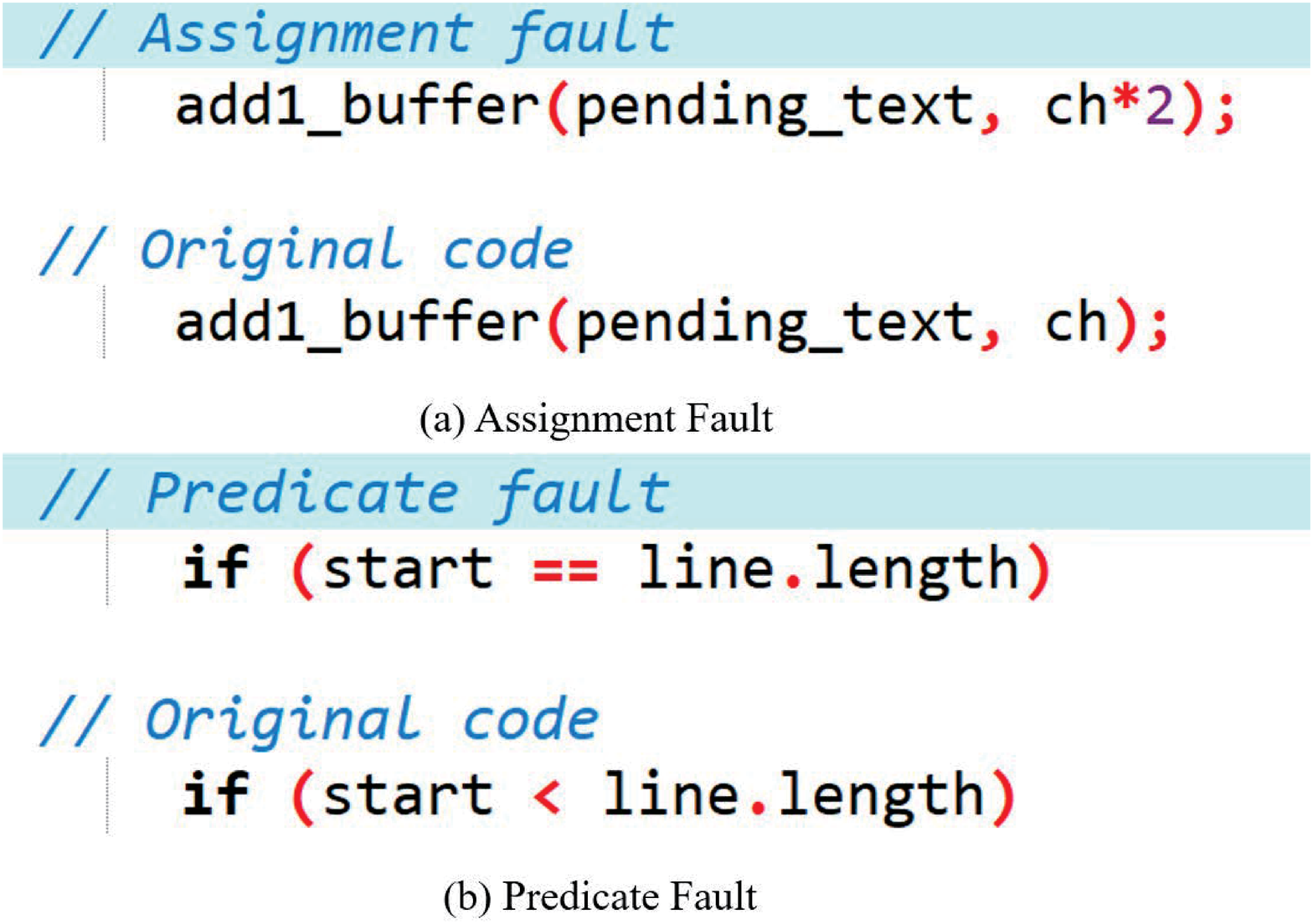}  
	\caption{Two fault types}
	\label{fig:faulttype}    
\end{figure}

\subsubsection{\synote{SIR programs}}
\label{subsubsect:sir}

\synote{SIR (Software-artifact Infrastructure Repository) contains a series of programs written in C that can be expropriated for the use of fault localization.} We employ mutation-based strategies to inject multiple artificial faults into four SIR benchmark programs for generating faulty versions~\cite{papadakis2019mutation[59]}. Research such as~\cite{andrews2005mutation[34], do2006use[35], liu2006statistical[36], andrews2006using[37], pradel2018deepbugs[78], just2014mutants[79]} has confirmed that mutation-based faults can simulate real-world faults and provide credible results for experiments in the field of software testing and debugging. The following two fault types are defined to mutate source code, which is exemplified in Figure \ref{fig:faulttype}:

\begin{itemize}
	\item $Assignment\ Fault$ \textbf{(AF)}: Editing a variable's value in the statement, or replacing the operators such as addition, subtraction, multiplication, division, etc. with each other (Figure \ref{fig:faulttype}(a));
	
	\item $Predicate\ Fault$ \textbf{(PF)}: Reversing the $if$-$else$ predicate, or deleting the $else$ statement, or modifying the decision condition, and so on. (Figure \ref{fig:faulttype}(b)).
\end{itemize}

After a mutation-based fault is seeded into a benchmark program, a 1-bug faulty version has been generated. To create an $r$-bug faulty version, the faults from $r$ individual 1-bug faulty versions are injected into the same program. This method of generating a multi-fault version by synthesizing multiple 1-bug faulty versions has been adopted by many studies~\cite{lamraoui2016formula[38], yu2015does[9], huang2013empirical[26]}.

A total of 960 multi-fault versions have been generated using 228 faults on SIR programs\footnote{When two or more specific faults exist in a program, the program may fail to compile, enter an infinite loop, or run for an excessive amount of time. These faulty versions were removed.}. From the perspective of NOF, they can be categorized into four classes, i.e., 2-bug, 3-bug, 4-bug, and 5-bug, according to how many faults a faulty version contains. On the other hand, from the perspective of FT, they can be categorized into three classes, i.e., TypeA, TypeP, and TypeH, according to the fault type(s) involved in a faulty version.

\begin{itemize}
	\item \textbf{TypeA}: This type of multi-fault version is generated by $r$ 1-bug faulty versions that contain assignment fault (each of $r$ faults contained in a TypeA faulty version is AF).
	
	\item \textbf{TypeP}: This type of multi-fault version is generated by $r$ 1-bug faulty versions that contain predicate fault (each of $r$ faults contained in a TypeP faulty version is PF).
	
	\item \textbf{TypeH}: This type of multi-fault version is generated by $r$ 1-bug faulty versions that contain both assignment fault and predicate fault (AF and PF are hybridly contained in a TypeH faulty version).
\end{itemize}

\subsubsection{\synote{Defects4J programs}}
\label{subsubsect:d4j}

\synote{Defects4J gathers a collection of real-world bugs from some open-source projects, due to the realism and ease-to-use, it has been becoming one of the most popular benchmarks in the current field of fault localization. Nonetheless, Defects4J is often utilized in single-fault rather than multi-fault environments, because each of its faulty versions only targets a specific fault. Recently, researchers revisited this benchmark and concluded a new point, that is, many of Defects4J faulty versions actually contain more than one fault, but only one of them can be revealed by the provided test suite. To adapt Defects4J to multi-fault scenarios, An et al. transplanted the fault-revealing test case(s) of another faulty version or other faulty versions to a basic faulty version, that is, enabling a strengthened test suite to detect more faults in the original program (i.e., the basic faulty version)~\cite{an2021searching[82]}.}

\synote{Following this strategy, a total of 100 multi-fault versions have been generated using 141 faults on Defects4J programs. It should be highlighted that the generation of multi-fault Defects4J programs involves two limitations. First, it is more difficult to generate multi-fault versions that contain more bugs. The faults in Defects4J come from real-world programming practice, to preserve such a characteristic, we use test cases transplantation instead of source code modification during the generation of multi-fault versions. Specifically, the majority of Defects4J faulty versions are indexed chronologically according to the revision date, a lower ID indicates a more recent version~\cite{an2021searching[82]}, thus the fault in a newer version is also likely to be contained in an older version. For example, we find that the fault in Lang-27 also appears in Lang-28, thus we can add the failed test case of Lang-27 to the test suite of Lang-28, for the generation of a 2-bug version, Lang-27-28. However, it is more difficult to search for a 5-bug version than a 2-bug version, since the more faults, the less likely they co-exist in a same program originally. For this reason, in the created 100 Defects4J multi-fault versions, half of them are 2-bug, and 25, 16, and 9 ones are 3-bug, 4-bug, and 5-bug, respectively. Second, as mentioned above, the faults in Defects4J are not obtained by artificial simulation, thus they cannot be properly categorized into assignment fault or predicate fault. As the consequence of these two problems, Defects4J programs are not suitable for exploring RQ2 (How NOF affects clustering effectiveness?) and RQ3 (Is clustering effectiveness affected by FT?).}

\synote{In summary, RQ1 and RQ4 will be investigated on all faulty versions that comprise both SIR and Defects4J, considering that these two topics do not involve the number of faults and fault types. And RQ2 and RQ3 will be investigated on SIR, since we can hardly set a proper and fair environment to explore the two questions on Defects4J.}

\subsection{Experiment setup}
\label{subsect3.2}

In this section, we elaborate on the experimental setups of the four RQs defined in Section \ref{sect1}.

\subsubsection{The risk evaluation formulas in SRR (RQ1)}
\label{subsubsect3.2.1}

Countless research has been conducted to investigate various REFs in the last four decades~\cite{wong2016survey[1], de2016spectrum[72]}. However, most of these studies proposed a novel REF or contrasted existing REFs empirically or theoretically in terms of its/their fault localization effectiveness, that is, analyzing the REF's capability to ranking the faulty statement(s) at the top of the list~\cite{naish2011model[39], xie2013theoretical[40], yoo2017human[73]}. 

For example, some novel REFs have emerged in the past ten years, including Crosstab~\cite{wong2011towards[18]} and DStar~\cite{wong2013dstar[44]} that were developed by Wong et al. in 2011 and 2013, respectively. The former constructs a crosstab for each statement in PUT to determine their suspiciousness by calculating the chi-square statistic and the coefficient of contingency, while the latter exponentially strengthens the function of $N_{CF}$ in spectrum information, making it more effective in fault localization than any other techniques compared with it according to the authors. Yoo created 30 novel REFs via genetic programming in 2012~\cite{yoo2012evolving[41]}, experimental results proved that GP-evolved REFs can consistently outperform many of the human-designed REFs. Xie et al. evaluated these 30 GP-evolved REFs using the theoretical framework in~\cite{xie2013theoretical[40]} and discovered three REFs with strong human competitiveness: GP02, GP03, and GP19~\cite{xie2013provably[42]}.

Apart from developing new REFs, some researchers have dedicated their effort to investigating a corpus of existing REFs. For example, Naish et al. investigated more than 30 REFs and extracted several equivalence relations guided by the strictest equivalence definition (i.e., only REFs that generate the same statement ranking lists are considered equivalent)~\cite{naish2011model[39]}. Xie et al. first excluded some REFs that are not intuitively justified in the context of SBFL, then selected 30 REFs from Naish et al.'s research to contrast them using a novel theoretical framework~\cite{xie2013theoretical[40]}. According to Naish et al. and Xie et al.'s conclusions, 30 REFs are divided into six equivalent groups that include 22 REFs and eight individual REFs.

To the best of our knowledge, no empirical study has been published to investigate how different REFs, which produce ranking lists that represent failed test cases, affect the clustering effectiveness in SRR-based parallel debugging. To fill this gap, we perform the first empirical study on the capability of 35 REFs in Table \ref{tab:35refs} to representing failed test cases.

\begin{table*}[]\footnotesize
	\begin{threeparttable}[b]
	\centering
	\setlength{\belowcaptionskip}{3pt}   
	\caption{\label{tab:35refs} 35 risk evaluation formulas} 
	\begin{tabular}{|l|c|l|c|}
		\hline
		\textbf{Name} & \textbf{Formula expression} & \textbf{Name} & \textbf{Formula expression} \\ \hline
		
		Naish1~\cite{naish2011model[39]}       & $\left\{\begin{array}{ll}
			-1 & \text { if } N_{C F}<N_{F} \\
			N_{S}-N_{C S} & \text { if } N_{C F}=N_{F}
		\end{array}\right.$           & Naish2~\cite{naish2011model[39]}        & $N_{C F} - \frac{N_{C S}}{N_{S} + 1}$  \\ \hline
	
	    Jaccard~\cite{chen2002pinpoint[51]} &  $\frac{N_{C F}}{N_{F} + N_{C S}}$  &
		Anderberg~\cite{naish2011model[39]}          &  $\frac{N_{C F}}{N_{C F} + 2(N_{U F} + N_{C S})}$   \\ \hline	
		
		Sørensen-Dice~\cite{naish2011model[39]}    & $\frac{2N_{C F}}{2N_{C F} + N_{U F} + N_{C S}}$ & Dice~\cite{naish2011model[39]}    &    $\frac{2N_{C F}}{N_{F} + N_{C S}}$ \\ \hline
		
		Goodman~\cite{naish2011model[39]}            &   $\frac{2N_{C F} - N_{U F} - N_{C S}}{2N_{C F} + N_{U F} + N_{C S}}$ & Tarantula~\cite{jones2005empirical[52]}          &     $\frac{\frac{N_{C F}}{N_{F}}}{\frac{N_{C F}}{N_{F}} + \frac{N_{C S}}{N_{S}}}$  \\ \hline
		
		qe~\cite{lee2009study[53]}                 &    $\frac{N_{C F}}{N_{C}}$ &
		CBI Inc.~\cite{liblit2005scalable[54]}           &      $\frac{N_{C F}}{N_{C}}$  - $\frac{N_{F}}{N}$   \\ \hline
		
		 Wong2~\cite{wong2007effective[55]}              &    $N_{C F} - N_{C S}$  &  Hamann~\cite{naish2011model[39]}             &    $\frac{N_{C F} + N_{U S} - N_{U F} - N_{C S}}{N}$ \\ \hline
		
		Simple Matching~\cite{naish2011model[39]}    &    $\frac{N_{C F} + N_{U S}}{N}$  & 	Sokal~\cite{naish2011model[39]}              &       $\frac{2(N_{C F} + N_{U S})}{2(N_{C F} + N_{U S}) + N_{U F} + N_{C S}}$ \\ \hline
		
		 Rogers \& Tanimoto~\cite{naish2011model[39]} &   $\frac{N_{C F} + N_{U S}}{N_{C F} + N_{U S} + 2(N_{U F} + N_{C S})}$  & Hamming etc.~\cite{naish2011model[39]}       &     $N_{C F} + N_{U S}$  \\ \hline
		 
		 Euclid~\cite{naish2011model[39]} & $\sqrt{N_{C F} + N_{U S}}$  &   Wong1~\cite{wong2007effective[55]}              &     $N_{C F}$ \\ \hline
		
		Russel \& Rao~\cite{naish2011model[39]}      &    $\frac{N_{C F}}{N}$  & 	Binary~\cite{naish2011model[39]}             &     $
		\left\{
		\begin{array}{ll}
			0  &  \textrm{if  $N_{C F}{<}N_{F}$}\\
			1  &  \textrm{if $N_{C F}{=}N_{F}$}
		\end{array}
		\right.
		$ \\ \hline
		
		Scott~\cite{naish2011model[39]}              &   $\frac{4N_{CF}N_{US}{-}4N_{UF}N_{CS}{-}(N_{UF}{-}N_{CS})^2}{(2N_{CF}{+}N_{UF}{+}N_{CS})(2N_{US}{+}N_{UF}{+}N_{CS})}$ & 	Rogot1~\cite{naish2011model[39]}             &  $\frac{1}{2}(\frac{N_{CF}}{2N_{CF}{+}N_{UF}{+}N_{CS}}{+}\frac{N_{US}}{2N_{US}{+}N_{UF}{+}N_{CS}})$  \\ \hline
		
		Kulczynski2~\cite{naish2011model[39]}        & $\frac{1}{2}(\frac{N_{CF}}{N_{F}}+\frac{N_{CF}}{N_{C}})$ &	Ochiai~\cite{abreu2006evaluation[31]}             &    $\frac{N_{CF}}{\sqrt{N_{F}N_{C}}}$  \\ \hline
		
		M2~\cite{naish2011model[39]}                 &   $\frac{N_{C F}} {N_{C F} + N_{U S} + 2(N_{U F}+ N_{C S})}$  & Ample2~\cite{naish2011model[39]}             &  $\frac{N_{CF}}{N_{F}}-\frac{N_{CS}}{N_{S}}$  \\ \hline
		
		Wong3~\cite{wong2007effective[55]}              &    $N_{C F}\ {-}\ h, \textrm{where }h{=}
		\left\{
		\begin{array}{ll}
			N_{C S}  &  \textrm{if  $N_{C S}{\leq}2$}\\
			2 {+} 0.1(N_{C S}{-}2)  &  \textrm{if $2{<}N_{C S}{\leq}10$}\\
			2.8 {+} 0.001(N_{C S}{-}10)  &  \textrm{if $N_{C S}{>}10$}
		\end{array}
		\right.
		$  & Arithmetic Mean~\cite{naish2011model[39]}    &     $\frac{2N_{CF}N_{US}{-}2N_{UF}N_{CS}}{N_{C}N_{U}{+}N_{F}N_{S}}$  \\ \hline
		
		 Cohen~\cite{naish2011model[39]}              &  $\frac{2N_{CF}N_{US}{-}2N_{UF}N_{CS}}{N_{C}N_{S}{+}N_{F}N_{U}}$ & 
		
		Fleiss~\cite{naish2011model[39]}             &   $\frac{4N_{CF}N_{US}{-}4N_{UF}N_{CS}{-}(N_{UF}{-}N_{CS})^2}{(2N_{CF}{+}N_{UF}{+}N_{CS}){+}(2N_{US}{+}N_{UF}{+}N_{CS})}$   \\ \hline
		
		Crosstab~\cite{wong2011towards[18]}  \tnote{*}
		& $\chi^{2}=\frac{\left(N_{C F}-E_{C F}\right)^{2}}{E_{C F}}+\frac{\left(N_{C S}-E_{C S}\right)^{2}}{E_{C S}}+\frac{\left(N_{U F}-E_{U F}\right)^{2}}{E_{U F}}+\frac{\left(N_{U S}-E_{U S}\right)^{2}}{E_{U S}}$ & 	DStar~\cite{wong2013dstar[44]} \tnote{**}           & $\frac{N_{C F}^*}{N_{U F} + N_{C S}}$ \\ \hline

	    GP02~\cite{yoo2012evolving[41]}               &  $2(N_{C F} + \sqrt{N_{U S}}) + \sqrt{N_{C S}}$  & 
		GP03~\cite{yoo2012evolving[41]}               &   $\sqrt{| N_{C F} ^ 2 - \sqrt{N_{C S}} |}$  \\ \hline
		
		GP19~\cite{yoo2012evolving[41]}               &   $N_{C F} \sqrt{|N_{C S} - N_{C F} + N_{U F} - N_{U S}|}$  & & \\ \hline		
	\end{tabular}
	\begin{tablenotes}
		\item[*] Crosstab will first calculate $\varphi$ for each statement to quantify its association with failed and successful executions, and then use $\varphi$ to determine if a statement should be assigned $\chi^2$, $-\chi^2$ or 0. Please refer to~\cite{wong2011towards[18]} for more details about this REF.
		\item[**] Considering the preference for DStar in many other studies (such as~\cite{pearson2017evaluating[47], arrieta2018spectrum[48]}), we set * = 2, the most thoroughly-explored value in our experiments.
	\end{tablenotes}
\end{threeparttable}
\end{table*}

\subsubsection{The number of faults in PUT (RQ2)}
\label{subsubsect3.2.2}

\iffalse
Although it can be intuitively inferred the more faults a program has, the more effort and time the debugging process will take, and despite the fact that a large body of research that explored the effect of the number of faults on fault localization effectiveness has been published~\cite{digiuseppe2011influence[4], digiuseppe2015fault[5], jones2002visualization[11]}, we still do not know how NOF affects the clustering step in parallel debugging. To that end, we observe and compare the effectiveness of clustering in 2-bug, 3-bug, 4-bug, and 5-bug scenarios.
\fi

\synote{The effect of the number of faults contained in a program on fault localization effectiveness has been investigated by many prior researchers~\cite{digiuseppe2011influence[4], digiuseppe2015fault[5], jones2002visualization[11]}, but how NOF affects the clustering stage in parallel debugging is still poorly explored. Although it is intuitive to assume that more bugs will lead to more failures, making it more difficult to divide them, we do not know whether this is reasonable from an empirical standpoint. To that purpose, we observe and compare the effectiveness of clustering in 2-bug, 3-bug, 4-bug, and 5-bug scenarios.}

\subsubsection{The fault type in PUT (RQ3)}
\label{subsubsect3.2.3}
Programmers may introduce various types of faults when coding due to unintentional mistakes or misunderstandings of programming logistics, as a result, FT is typically unpredictable because of the randomness and uncertainty of onsite programming. Lamraoui and Nakajima categorized common faults in multi-fault scenarios into several types, including data-flow dependent faults and control-dependent faults~\cite{lamraoui2016formula[38]}. Similar to these, we define assignment faults and predicate faults, two types of faults that are most likely to occur in programming as our research objects, and accordingly generate a series of TypeA faulty versions with only assignment faults, TypeP faulty versions with only predicate faults, and TypeH faulty versions with both two types of faults to observe clustering effectiveness.

\subsubsection{The number of successful test cases paired with one individual failed test case (RQ4)}
\label{subsubsect3.2.4}
While clustering failed test cases via SRR, many prior studies paired one failed test case with all successful test cases and input them into an REF to produce a ranking list representing this failed test case, without explaining why $all$ successful test cases are employed here. In fact, many studies including~\cite{sun2016ipsetful[61], mottaghi2017test[62]} have managed to utilize test case selection or test suite reduction techniques to lower debugging expenses, some recent studies have also investigated the impact of test suites on fault localization~\cite{lei2018test[65], perez2017test[66]}. For example, as Fu et al. argued, if the number of successful test cases is too large, the noise will be introduced into the fault localization process~\cite{fu2017test[63]}.  However, these works only evaluated the effect of the number of test cases on fault localization, not fault isolation built upon SRR. We try to cut the scale of successful test cases utilized in SRR by pairing 100\%, 80\%, 60\%, 40\%, and 20\% of successful test cases with one failed test case,  respectively, to monitor if the clustering effectiveness declines as NSP1F falls.

\subsection{Metrics}
\label{subsect3.3}

Two classes of metrics, external metrics~\cite{wu2009adapting[74]} and internal metrics~\cite{tan2016introduction[75]}, are typically implemented to measure the effectiveness of clustering techniques. The former contrast clustering results with the oracle, while the latter examine inherent properties of clustering results, such as compactness and separation, without using an off-the-shelf baseline~\cite{xie2017new[60]}. While clustering failed test cases in parallel debugging, ideal outputs should exhibit linkages between each failed test case in TS and each fault in PUT, which is available in our controlled experiments. Therefore, we employ four widely-used external metrics, JC, FMI, PR and RR, to evaluate the experimental results.

\subsubsection{Pair of cases-based metric}
\label{subsubsect3.3.1}

The pair of cases-based metric refers to compare the indexing consistency of each pair of failed test cases in the generated cluster with the oracle cluster. Four scenarios in which are depicted in Table \ref{tab:pairmetric}.

\begin{table}[]\footnotesize
	\centering
	\setlength{\belowcaptionskip}{3pt}
	\caption{\label{tab:pairmetric} Four scenarios in the pair of cases-based metric} 
	\begin{tabular}{cll}
		\hline
		\multirow{2}{*}{\textbf{Notation}} & \multicolumn{2}{c}{\textbf{Results of failure indexing}}                           \\ \cline{2-3}  & 
        In the generated cluster & In the oracle cluster \\ \hline
		SS                       & Same                            & Same                         \\
		SD                      & Same                            & Difference              \\
		DS                      & Difference                  & Same                         \\
		DD                      & Difference                 & Difference       \\ \hline         
	\end{tabular}
\end{table}

Assuming that there are $n$ failed test cases that need to be clustered, a total of $C_n^2$ pairs will be examined in the pair of cases-based metric. The numbers of pairs that fall into SS, SD, DS, and DD categories are denoted as $X_{SS}$, $X_{SD}$, $X_{DS}$, and $X_{DD}$, respectively.

The above notations can be incorporated into the Jaccard Coefficient (JC) and the Fowlkes and Mallows Index (FMI), which are defined in Formula \ref{equ4} and Formula \ref{equ5}, respectively. JC and FMI are used to determine the similarity between the generated cluster and the oracle cluster, for measuring the clustering results~\cite{huangexploration[43]}.

\begin{equation}
	\label{equ4}
	JC = \frac{X_{SS}}{X_{SS} + X_{SD} + X_{DS}}
\end{equation}

\vspace{6pt}

\begin{equation}
	\label{equ5}
	FMI = \sqrt{\frac{X_{SS}}{X_{SS} + X_{SD}} \times \frac{X_{SS}}{X_{SS} + X_{DS}}}
\end{equation}

It can be proved that the intervals of JC and FMI are both [0, 1], and that the larger the value in this range, the more effective clustering is. A simple example is given below to describe JC and FMI.

\begin{figure}
	\centering  
	\includegraphics[height=2.2cm,width=5cm]{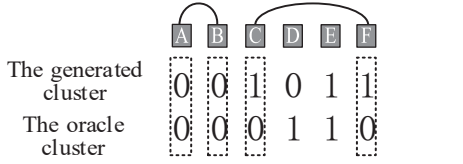}  
	\caption{SS pairs in the generated cluster and the oracle cluster}
	\label{fig:sspairs}    
\end{figure}

As shown in Figure \ref{fig:sspairs}, six failed test cases ($A$, $B$, $C$, $D$, $E$, and $F$) are indexed divergently in the generated cluster and the oracle cluster. Among the $C_6^2$ = 15 pairs of cases ($A$-$B$, $A$-$C$, $A$-$D$, ···, $E$-$F$), $A$-$B$ and $C$-$F$ are in the $same$ cluster in the generated cluster, and also in the $same$ cluster in the oracle cluster, which meets the scenario SS in Table \ref{tab:pairmetric}, therefore, $X_{SS}$ = 2. Similarly, we can get $X_{SD}$ = 4, $X_{DS}$ = 5, and $X_{DD}$ = 4. Incorporating these notations into Formulas \ref{equ4} and Formula \ref{equ5}, JC and FMI will be set to 0.182 and 0.309, respectively.

\subsubsection{Single case-based metric}
\label{subsubsect3.3.2}

The single case-based metric refers to compare the classification result of each failed test case in the generated cluster with the oracle cluster. Four scenarios in which are depicted in Table \ref{tab:singlemetric}.

\begin{table}[]\footnotesize
	\centering
	\setlength{\belowcaptionskip}{3pt}
	\caption{\label{tab:singlemetric} Four scenarios in the single case-based metric} 
	\begin{tabular}{cll}
		\hline
		\multirow{2}{*}{\textbf{Notation}} & \multicolumn{2}{c}{\textbf{Results of failure indexing}}                           \\ \cline{2-3}  & 
		In the generated cluster & In the oracle cluster \\ \hline
		TP                       & Positive                     & Positive                         \\
		FP                       & Positive                     & Negative               \\
		TN                      & Negative                   & Negative                         \\
		FN                      &Negative                    & Positive        \\ \hline         
	\end{tabular}
\end{table}

The numbers of failed test cases that fall into TP, FP, TN, and FN categories are denoted as $X_{TP}$, $X_{FP}$, $X_{TN}$, and $X_{FN}$, respectively. 

The above notations can be incorporated into the Precision Rate (PR) and the Recall Rate (RR), which are defined in Formula \ref{equ6} and Formula \ref{equ7}, respectively, for measuring the clustering results.

\begin{equation}
	\label{equ6}
	PR = \frac{X_{TP}}{X_{TP} + X_{FP}}
\end{equation}

\begin{equation}
	\label{equ7}
	RR = \frac{X_{TP}}{X_{TP} + X_{FN}}
\end{equation}

It can be proved that the intervals of PR and RR are both [0, 1], and that the larger the value in this range, the more effective clustering is.

\synote{As shown in Figure \ref{fig:sspairs}, failed test cases $D$ and $E$ are labelled as positive, and the remaining four ones are labelled as negative in the oracle cluster. But in the generated cluster, failed test cases $C$ and $F$ are wrongly labelled as positive, thus the value of $X_{FP}$ can be determined as 2. Similarly, we can get $X_{TP}$ = 1, $X_{TN}$ = 2, and $X_{FN}$ = 1. Incorporating these notations into Formulas \ref{equ6} and Formula \ref{equ7}, PR and RR will be set to 0.333 and 0.5, respectively.}

\subsubsection{The virtual mapping problem}
\label{subsubsect3.3.3}

It should be noted that the different permutations between generated clusters and oracle clusters will result in different outputs of the external metrics, and the diversity of permutations will significantly grow with the number of faults increases.  For example, in a 2-bug scenario, the permutations between two generated clusters and two oracle clusters are $A_2^2$ = 2, while in a 5-bug scenario, the permutations between five generated clusters and five oracle clusters are $A_5^5$ = 120. Such diversity of permutations does not exist in practical parallel debugging, it only occurs in the contrast between the output and the oracle. In other words, each developer will be allocated to a fault-focused TS and will be responsible for localizing the corresponding fault independently, thus regardless of how many potential permutations exist, there is \textbf{only one} real combination of generated clusters and oracle clusters. As a result, the $permutation$ between generated clusters and oracle clusters in experiments is the $combination$ in practice (we call this problem the $virtual\ mapping\ problem$). In our experiments, we extract faulty versions in which the number of faults has been precisely estimated, i.e., the number of faults equals the number of generated clusters, to perform analyses. For each of these faulty versions, we enumerate all feasible permutations followed by picking the optimal one based on the value of JC, FMI, PR, or RR for evaluation, because which permutation reflects the real mapping relations is unknown.

\section{RESULT AND ANALYSIS}
\label{sect4}

We conduct extensive controlled experiments according to the research questions in Section \ref{sect1} and predesigned setups in Section \ref{sect3}. Experimental results and analyses are given in this section. 

\subsection{The capability of different REFs to representing failed test cases (RQ1)}
\label{subsect4.1}

We reorganize 35 REFs in Table \ref{tab:35refs} into 12 disjoint groups, as shown in Table \ref{tab:12groups}, because we find that some REFs have the same performance in representing failed test cases (details are omitted to conserve space). Only one REF (in bold) in each group is selected for analyses since its capability to representing failed test cases is equal to the others in the group it belongs to.

\begin{table}[]\footnotesize
	\centering
	\setlength{\belowcaptionskip}{3pt}
	\caption{\label{tab:12groups} 12 groups of risk evaluation formulas with the same capability to representing failed test cases} 
	\begin{tabular}{|c|p{6cm}|}
		\hline
		\textbf{Name}    & \multicolumn{1}{c|}{\textbf{REFs}}                                                        \\ \hline
		\textbf{Group1}  & \textbf{Naish2}                                                                           \\ \hline
		\textbf{Group2}  & \textbf{Jaccard}, Anderberg, Sørensen-Dice, Dice, Goodman, M2, Naish1, DStar              \\ \hline
		\textbf{Group3}  & \textbf{Tarantula}, qe, CBI Inc, Kulczynski2, Ochiai                                      \\ \hline
		\textbf{Group4}  & \textbf{Wong2}, Hamann, Simple Matching, Sokal, Rogers \& Tanimoto, Hamming etc., Euclid \\ \hline
		\textbf{Group5}  & \textbf{Wong1}, Binary, Russel \& Rao                                                   \\ \hline
		\textbf{Group6}  & \textbf{Scott}, Rogot1                                                                    \\ \hline
		\textbf{Group7}  & \textbf{Ample2}, Arithmetic Mean, Cohen, Crosstab                                         \\ \hline
		\textbf{Group8}  & \textbf{Wong3}                                                                            \\ \hline
		\textbf{Group9}  & \textbf{Fleiss}                                                                           \\ \hline
		\textbf{Group10} & \textbf{GP02}                                                                             \\ \hline
		\textbf{Group11} & \textbf{GP03}                                                                             \\ \hline
		\textbf{Group12} & \textbf{GP19}                                                                             \\ \hline
	\end{tabular}
\end{table}

\begin{table*}[]\footnotesize
	\centering
	\setlength{\belowcaptionskip}{3pt}
	\caption{\label{tab:contrast} \synote{Contrast of the capability of 12 groups of REFs to representing failed test cases}} 
	\resizebox{\textwidth}{!}{
		\begin{tabular}{lllllllllllll}
			\hline
			\diagbox{\ $R_2$}{Versus} {\ $R_1$}   & \textbf{Group1}                      & \textbf{Group2}                      & \textbf{Group3}                     & \textbf{Group4} & \textbf{Group5}                      & \textbf{Group6}                      & \textbf{Group7}                      & \textbf{Group8}                      & \textbf{Group9}                      & \textbf{Group10}                     & \textbf{Group11}                     & \textbf{Group12}                     \\ \hline
			\textbf{Group1}  & \textbf{}                            & {\color[HTML]{2D6AA7} $\checkmark$}{\color[HTML]{5AC5C6} $\checkmark$}{\color[HTML]{44972B} $\checkmark$}{\color[HTML]{F19E37} $\checkmark$}                                & \textbf{}                           & \textbf{}       & {\color[HTML]{2D6AA7} $\checkmark$}{\color[HTML]{5AC5C6} $\checkmark$}{\color[HTML]{44972B} $\checkmark$}{\color[HTML]{F19E37} $\checkmark$} & {\color[HTML]{2D6AA7} $\checkmark$}{\color[HTML]{5AC5C6} $\checkmark$}{\color[HTML]{44972B} $\checkmark$}{\color[HTML]{F19E37} $\checkmark$} & \textbf{}                            & \textbf{}                            & {\color[HTML]{2D6AA7} $\checkmark$}{\color[HTML]{5AC5C6} $\checkmark$}{\color[HTML]{44972B} $\checkmark$}{\color[HTML]{F19E37} $\checkmark$} & \textbf{}                            & {\color[HTML]{2D6AA7} $\checkmark$}{\color[HTML]{5AC5C6} $\checkmark$}{\color[HTML]{44972B} $\checkmark$}{\color[HTML]{F19E37} $\checkmark$} & {\color[HTML]{2D6AA7} $\checkmark$}{\color[HTML]{5AC5C6} $\checkmark$}{\color[HTML]{44972B} $\checkmark$}{\color[HTML]{F19E37} $\checkmark$} \\
			\textbf{Group2}  & \textbf{}                            & \textbf{}                            & \textbf{}                           & \textbf{}       & {\color[HTML]{2D6AA7} $\checkmark$}{\color[HTML]{5AC5C6} $\checkmark$}{\color[HTML]{44972B} $\checkmark$}{\color[HTML]{F19E37} $\checkmark$} & {\color[HTML]{2D6AA7} $\checkmark$}{\color[HTML]{5AC5C6} $\checkmark$}{\color[HTML]{44972B} $\checkmark$}{\color[HTML]{F19E37} $\checkmark$} & \textbf{}                            & \textbf{}                            & {\color[HTML]{2D6AA7} $\checkmark$}{\color[HTML]{5AC5C6} $\checkmark$}{\color[HTML]{44972B} $\checkmark$}{\color[HTML]{F19E37} $\checkmark$} & \textbf{}                            & {\color[HTML]{2D6AA7} $\checkmark$}{\color[HTML]{5AC5C6} $\checkmark$}{\color[HTML]{44972B} $\checkmark$}{\color[HTML]{F19E37} $\checkmark$} & {\color[HTML]{2D6AA7} $\checkmark$}{\color[HTML]{5AC5C6} $\checkmark$}{\color[HTML]{44972B} $\checkmark$}{\color[HTML]{F19E37} $\checkmark$} \\
			\textbf{Group3}  & {\color[HTML]{2D6AA7} $\checkmark$}{\color[HTML]{5AC5C6} $\checkmark$}{\color[HTML]{44972B} $\checkmark$}{\color[HTML]{F19E37} $\checkmark$} & {\color[HTML]{2D6AA7} $\checkmark$}{\color[HTML]{5AC5C6} $\checkmark$}{\color[HTML]{44972B} $\checkmark$}{\color[HTML]{F19E37} $\checkmark$} & \textbf{}                           & \textbf{}       & {\color[HTML]{2D6AA7} $\checkmark$}{\color[HTML]{5AC5C6} $\checkmark$}{\color[HTML]{44972B} $\checkmark$}{\color[HTML]{F19E37} $\checkmark$} & {\color[HTML]{2D6AA7} $\checkmark$}{\color[HTML]{5AC5C6} $\checkmark$}{\color[HTML]{44972B} $\checkmark$}{\color[HTML]{F19E37} $\checkmark$} & {\color[HTML]{2D6AA7} $\checkmark$}{\color[HTML]{5AC5C6} $\checkmark$}{\color[HTML]{44972B} $\checkmark$}{\color[HTML]{F19E37} $\checkmark$} & {\color[HTML]{2D6AA7} $\checkmark$}{\color[HTML]{5AC5C6} $\checkmark$}{\color[HTML]{44972B} $\checkmark$}  & {\color[HTML]{2D6AA7} $\checkmark$}{\color[HTML]{5AC5C6} $\checkmark$}{\color[HTML]{44972B} $\checkmark$}{\color[HTML]{F19E37} $\checkmark$} & {\color[HTML]{2D6AA7} $\checkmark$}{\color[HTML]{5AC5C6} $\checkmark$}{\color[HTML]{44972B} $\checkmark$}{\color[HTML]{F19E37} $\checkmark$} & {\color[HTML]{2D6AA7} $\checkmark$}{\color[HTML]{5AC5C6} $\checkmark$}{\color[HTML]{44972B} $\checkmark$}{\color[HTML]{F19E37} $\checkmark$} & {\color[HTML]{2D6AA7} $\checkmark$}{\color[HTML]{5AC5C6} $\checkmark$}{\color[HTML]{44972B} $\checkmark$}{\color[HTML]{F19E37} $\checkmark$} \\
			\textbf{Group4}  & {\color[HTML]{2D6AA7} $\checkmark$}{\color[HTML]{5AC5C6} $\checkmark$}{\color[HTML]{44972B} $\checkmark$}{\color[HTML]{F19E37} $\checkmark$} & {\color[HTML]{2D6AA7} $\checkmark$}{\color[HTML]{5AC5C6} $\checkmark$}{\color[HTML]{44972B} $\checkmark$}{\color[HTML]{F19E37} $\checkmark$} & {\color[HTML]{2D6AA7} $\checkmark$}{\color[HTML]{5AC5C6} $\checkmark$}{\color[HTML]{44972B} $\checkmark$}{\color[HTML]{F19E37} $\checkmark$}                                & \textbf{}       & {\color[HTML]{2D6AA7} $\checkmark$}{\color[HTML]{5AC5C6} $\checkmark$}{\color[HTML]{44972B} $\checkmark$}{\color[HTML]{F19E37} $\checkmark$} & {\color[HTML]{2D6AA7} $\checkmark$}{\color[HTML]{5AC5C6} $\checkmark$}{\color[HTML]{44972B} $\checkmark$}{\color[HTML]{F19E37} $\checkmark$} & {\color[HTML]{2D6AA7} $\checkmark$}{\color[HTML]{5AC5C6} $\checkmark$}{\color[HTML]{44972B} $\checkmark$}{\color[HTML]{F19E37} $\checkmark$} & {\color[HTML]{2D6AA7} $\checkmark$}{\color[HTML]{5AC5C6} $\checkmark$}{\color[HTML]{44972B} $\checkmark$}{\color[HTML]{F19E37} $\checkmark$} & {\color[HTML]{2D6AA7} $\checkmark$}{\color[HTML]{5AC5C6} $\checkmark$}{\color[HTML]{44972B} $\checkmark$}{\color[HTML]{F19E37} $\checkmark$} & {\color[HTML]{2D6AA7} $\checkmark$}{\color[HTML]{5AC5C6} $\checkmark$}{\color[HTML]{44972B} $\checkmark$}{\color[HTML]{F19E37} $\checkmark$} & {\color[HTML]{2D6AA7} $\checkmark$}{\color[HTML]{5AC5C6} $\checkmark$}{\color[HTML]{44972B} $\checkmark$}{\color[HTML]{F19E37} $\checkmark$} & {\color[HTML]{2D6AA7} $\checkmark$}{\color[HTML]{5AC5C6} $\checkmark$}{\color[HTML]{44972B} $\checkmark$}{\color[HTML]{F19E37} $\checkmark$} \\
			\textbf{Group5}  & \textbf{}                            & \textbf{}                            & \textbf{}                           & \textbf{}       & \textbf{}                            & \textbf{}                            & \textbf{}                            & \textbf{}                            & $\  \ \ \ \ \ $ {\color[HTML]{44972B} $\checkmark$}                            & \textbf{}                            & {\color[HTML]{2D6AA7} $\checkmark$} $\ $ {\color[HTML]{44972B} $\checkmark$}{\color[HTML]{F19E37} $\checkmark$} & {\color[HTML]{2D6AA7} $\checkmark$}{\color[HTML]{5AC5C6} $\checkmark$}{\color[HTML]{44972B} $\checkmark$}{\color[HTML]{F19E37} $\checkmark$} \\
			\textbf{Group6}  & \textbf{}                            & \textbf{}                            & \textbf{}                           & \textbf{}       & {\color[HTML]{2D6AA7} $\checkmark$}{\color[HTML]{5AC5C6} $\checkmark$}{\color[HTML]{44972B} $\checkmark$}{\color[HTML]{F19E37} $\checkmark$} & \textbf{}                            & \textbf{}                            & \textbf{}                            & {\color[HTML]{2D6AA7} $\checkmark$}{\color[HTML]{5AC5C6} $\checkmark$}{\color[HTML]{44972B} $\checkmark$}{\color[HTML]{F19E37} $\checkmark$} & \textbf{}                            & {\color[HTML]{2D6AA7} $\checkmark$}{\color[HTML]{5AC5C6} $\checkmark$}{\color[HTML]{44972B} $\checkmark$}{\color[HTML]{F19E37} $\checkmark$} & {\color[HTML]{2D6AA7} $\checkmark$}{\color[HTML]{5AC5C6} $\checkmark$}{\color[HTML]{44972B} $\checkmark$}{\color[HTML]{F19E37} $\checkmark$} \\
			\textbf{Group7}  & {\color[HTML]{2D6AA7} $\checkmark$}{\color[HTML]{5AC5C6} $\checkmark$}{\color[HTML]{44972B} $\checkmark$}{\color[HTML]{F19E37} $\checkmark$} & {\color[HTML]{2D6AA7} $\checkmark$}{\color[HTML]{5AC5C6} $\checkmark$}{\color[HTML]{44972B} $\checkmark$}{\color[HTML]{F19E37} $\checkmark$} & \textbf{}                           & \textbf{}       & {\color[HTML]{2D6AA7} $\checkmark$}{\color[HTML]{5AC5C6} $\checkmark$}{\color[HTML]{44972B} $\checkmark$}{\color[HTML]{F19E37} $\checkmark$} & {\color[HTML]{2D6AA7} $\checkmark$}{\color[HTML]{5AC5C6} $\checkmark$}{\color[HTML]{44972B} $\checkmark$}{\color[HTML]{F19E37} $\checkmark$} & \textbf{}                            & \textbf{}                            & {\color[HTML]{2D6AA7} $\checkmark$}{\color[HTML]{5AC5C6} $\checkmark$}{\color[HTML]{44972B} $\checkmark$}{\color[HTML]{F19E37} $\checkmark$} & \textbf{}                            & {\color[HTML]{2D6AA7} $\checkmark$}{\color[HTML]{5AC5C6} $\checkmark$}{\color[HTML]{44972B} $\checkmark$}{\color[HTML]{F19E37} $\checkmark$} & {\color[HTML]{2D6AA7} $\checkmark$}{\color[HTML]{5AC5C6} $\checkmark$}{\color[HTML]{44972B} $\checkmark$}{\color[HTML]{F19E37} $\checkmark$} \\
			\textbf{Group8}  & {\color[HTML]{2D6AA7} $\checkmark$}{\color[HTML]{5AC5C6} $\checkmark$}{\color[HTML]{44972B} $\checkmark$}{\color[HTML]{F19E37} $\checkmark$} & {\color[HTML]{2D6AA7} $\checkmark$}{\color[HTML]{5AC5C6} $\checkmark$}{\color[HTML]{44972B} $\checkmark$}{\color[HTML]{F19E37} $\checkmark$} & $\qquad \ \ ${\color[HTML]{F19E37} $\checkmark$}      & \textbf{}       & {\color[HTML]{2D6AA7} $\checkmark$}{\color[HTML]{5AC5C6} $\checkmark$}{\color[HTML]{44972B} $\checkmark$}{\color[HTML]{F19E37} $\checkmark$} & {\color[HTML]{2D6AA7} $\checkmark$}{\color[HTML]{5AC5C6} $\checkmark$}{\color[HTML]{44972B} $\checkmark$}{\color[HTML]{F19E37} $\checkmark$} & {\color[HTML]{2D6AA7} $\checkmark$}{\color[HTML]{5AC5C6} $\checkmark$}{\color[HTML]{44972B} $\checkmark$}{\color[HTML]{F19E37} $\checkmark$} & \textbf{}                            & {\color[HTML]{2D6AA7} $\checkmark$}{\color[HTML]{5AC5C6} $\checkmark$}{\color[HTML]{44972B} $\checkmark$}{\color[HTML]{F19E37} $\checkmark$} & {\color[HTML]{2D6AA7} $\checkmark$}{\color[HTML]{5AC5C6} $\checkmark$}{\color[HTML]{44972B} $\checkmark$}{\color[HTML]{F19E37} $\checkmark$}                           & {\color[HTML]{2D6AA7} $\checkmark$}{\color[HTML]{5AC5C6} $\checkmark$}{\color[HTML]{44972B} $\checkmark$}{\color[HTML]{F19E37} $\checkmark$} & {\color[HTML]{2D6AA7} $\checkmark$}{\color[HTML]{5AC5C6} $\checkmark$}{\color[HTML]{44972B} $\checkmark$}{\color[HTML]{F19E37} $\checkmark$} \\
			\textbf{Group9}  & \textbf{}                            & \textbf{}                            & \textbf{}                           & \textbf{}       & {\color[HTML]{2D6AA7} $\checkmark$}{\color[HTML]{5AC5C6} $\checkmark$} $\ $ {\color[HTML]{F19E37} $\checkmark$} & \textbf{}                            & \textbf{}                            & \textbf{}                            & \textbf{}                            & \textbf{}                            & {\color[HTML]{2D6AA7} $\checkmark$}{\color[HTML]{5AC5C6} $\checkmark$}{\color[HTML]{44972B} $\checkmark$}{\color[HTML]{F19E37} $\checkmark$} & {\color[HTML]{2D6AA7} $\checkmark$}{\color[HTML]{5AC5C6} $\checkmark$}{\color[HTML]{44972B} $\checkmark$}{\color[HTML]{F19E37} $\checkmark$} \\
			\textbf{Group10} & {\color[HTML]{2D6AA7} $\checkmark$}{\color[HTML]{5AC5C6} $\checkmark$}{\color[HTML]{44972B} $\checkmark$}{\color[HTML]{F19E37} $\checkmark$} & {\color[HTML]{2D6AA7} $\checkmark$}{\color[HTML]{5AC5C6} $\checkmark$}{\color[HTML]{44972B} $\checkmark$}{\color[HTML]{F19E37} $\checkmark$} &  \textbf{} & \textbf{}       & {\color[HTML]{2D6AA7} $\checkmark$}{\color[HTML]{5AC5C6} $\checkmark$}{\color[HTML]{44972B} $\checkmark$}{\color[HTML]{F19E37} $\checkmark$} & {\color[HTML]{2D6AA7} $\checkmark$}{\color[HTML]{5AC5C6} $\checkmark$}{\color[HTML]{44972B} $\checkmark$}{\color[HTML]{F19E37} $\checkmark$} & {\color[HTML]{2D6AA7} $\checkmark$}{\color[HTML]{5AC5C6} $\checkmark$}{\color[HTML]{44972B} $\checkmark$}{\color[HTML]{F19E37} $\checkmark$} & \textbf{} & {\color[HTML]{2D6AA7} $\checkmark$}{\color[HTML]{5AC5C6} $\checkmark$}{\color[HTML]{44972B} $\checkmark$}{\color[HTML]{F19E37} $\checkmark$} & \textbf{}                            & {\color[HTML]{2D6AA7} $\checkmark$}{\color[HTML]{5AC5C6} $\checkmark$}{\color[HTML]{44972B} $\checkmark$}{\color[HTML]{F19E37} $\checkmark$} & {\color[HTML]{2D6AA7} $\checkmark$}{\color[HTML]{5AC5C6} $\checkmark$}{\color[HTML]{44972B} $\checkmark$}{\color[HTML]{F19E37} $\checkmark$} \\
			\textbf{Group11} & \textbf{}                            & \textbf{}                            & \textbf{}                           & \textbf{}       & $\ \ $ {\color[HTML]{5AC5C6} $\checkmark$}                           & \textbf{}                            & \textbf{}                            & \textbf{}                            & \textbf{}                            & \textbf{}                            & \textbf{}                            & {\color[HTML]{2D6AA7} $\checkmark$}{\color[HTML]{5AC5C6} $\checkmark$}{\color[HTML]{44972B} $\checkmark$}{\color[HTML]{F19E37} $\checkmark$} \\
			\textbf{Group12} & \textbf{}                            & \textbf{}                            & \textbf{}                           & \textbf{}       & \textbf{}                            & \textbf{}                            & \textbf{}                            & \textbf{}                            & \textbf{}                            & \textbf{}                            & \textbf{}                            & \textbf{}                            \\ \hline
		\end{tabular}}

\end{table*}

For each of faulty versions, we implement the workflow shown in Figure \ref{fig:workflow} to estimate the number of clusters based on the ranking lists produced by an REF. There are three scenarios in this stage:

\begin{itemize}
	\item\textbf{Under:} The estimated number of clusters is fewer than NOF (i.e., $k$ $<$ $r$).
	\item\textbf{Equal:}  The estimated number of clusters is equal to NOF (i.e., $k$ == $r$).
	\item\textbf{Over:} The estimated number of clusters exceeds NOF (i.e., $k$ $>$ $r$).
\end{itemize}

\begin{figure}  
	\centering  
	\includegraphics[height=5.4cm,width=7.6cm]{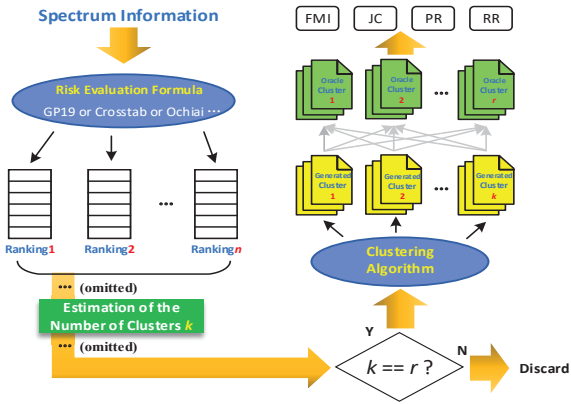}  
	\caption{The virtual mapping process (with checking whether the estimated number of clusters equals NOF)}  
	\label{fig:virtual}  
\end{figure}

If the estimated number of clusters $k$  in a faulty version is equal to NOF $r$, we send this faulty version to the next clustering step. Otherwise, if $k$ is not equal to $r$, this faulty version is discarded. \synote{In a real multi-fault localization scenario, even if the estimated number of clusters $k$ and the number of faults $r$ are not identical, the whole process can also be continued: if $k > r$, localization can be stopped when all failures disappear, and if $k < r$, localization can be carried out more than one iteration. This paper focuses on clustering rather than the following localization stage, the evaluation of clustering effectiveness is the main purpose, thus we do not take ``$k \neq r$'' scenarios into account.} The filtering, as well as the follow-up virtual mapping process, are illustrated in Figure \ref{fig:virtual}.

\begin{figure}   
	\includegraphics[height=6.3cm,width=8.5cm]{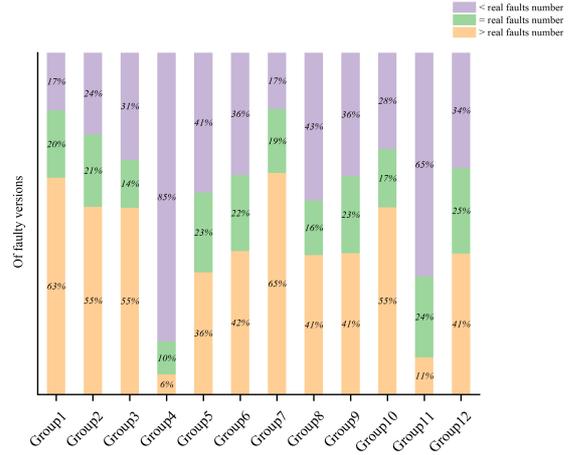}  
	\caption{\synote{$V_{under}^R$, $V_{equal}^R$, and $V_{over}^R$ of 12 groups of REFs}}  
	\label{fig:equalversion}  
\end{figure}

When estimating the number of clusters based on the ranking lists produced by an REF $R$, we denote the numbers of faulty versions that fall into the $Under$, $Equal$, and $Over$ categories as $V_{under}^R$, $V_{equal}^R$, and $V_{over}^R$, respectively. For an REF $R$, a greater $V_{equal}^R$, as well as a fewer $V_{under}^R$ and a fewer $V_{over}^R$, partly indicate that $R$ captures the execution features of failed test cases more effectively thus can better represent them. $V_{under}^R$, $V_{equal}^R$, and $V_{over}^R$ of each group of REFs on all faulty versions are shown in Figure \ref{fig:equalversion}  \footnote{The longer the $green$ band, the more faulty versions’ NOF can be accurately estimated based on the ranking lists produced by the corresponding REF.}.

It can be seen that based on the ranking lists produced by Group12, NOF is accurately estimated on \synote{25}\% of the \synote{1060} faulty versions  ($V_{equal}^{Group12}$ = \synote{265}). Besides, based on the ranking lists produced by Group7, the estimated numbers of clusters on \synote{65}\% of faulty versions exceed the NOF ($V_{over}^{Group7}$ = \synote{687}), implying this REF is $over$-$representing$ in modeling failed test cases (i.e., too sensitive to model failures to a nicety). Based on the ranking lists produced by Group4, the estimated numbers of clusters in \synote{85}\% of faulty versions are fewer than the NOF ($V_{under}^{Group4}$ = \synote{896}), indicating that this REF appears $under$-$representing$ in modeling failed test cases (i.e., too deficient to model failures distinguishably).

We select only faulty versions that fall into the $Equal$ category (i.e., satisfy the ``$k$ == $r$" criteria) for clustering, as illustrated in Figure \ref{fig:virtual}. The capability of an REF $R$ to representing failed test cases can be assessed by two indicators: the value of $V_{equal}^R$ and the clustering effectiveness on these $V_{equal}^R$  faulty versions. We define the ${Sum\_Metric}_{M}^{R}$, as shown in Formula \ref{equ8}, to incorporate the two metrics into a single value.

\begin{equation}
	\label{equ8}
	{Sum\_Metric}_{M}^{R}=\sum_{i}^{V_{equal}^{R}} M_{i}
\end{equation}

Where $R$ represents the REF, $M_i$ is the value of the clustering metric $M$ ($M$ takes FMI, JC, PR or RR) on the $i^{\rm{th}}$ faulty version. \synote{For instance, if we want to evaluate Group12 from the standpoint of FMI (i.e., $R$ takes Group12 and $M$ takes FMI) using Formula \ref{equ8}, we can first get the value of $V_{equal}^{Group12}$ (265), and then calculate $Sum\_Metric_{FMI}^{Group12}$ by adding up $FMI_{1}$, $FMI_{2}$, ..., $FMI_{265}$.} \minorrevised{Specifically, the values of $FMI$ on the first, the second, ..., and the $265^{\rm{th}}$ version are 0.78, 0.77, ..., and 1.00, respectively, thus the value of  $Sum\_Metric_{FMI}^{Group12}$ can be determined by adding up 0.78, 0.77, ..., and 1.00, that is, 221.33.} Obviously, the greater the $V_{equal}^R$ value of the REF $R$, the more possibility it has to obtain a greater ${Sum\_Metric}_{M}^{R}$.

\begin{table}[]\footnotesize
	\centering
	\setlength{\belowcaptionskip}{3pt}
	\caption{\label{tab:summetric} \synote{The values of $Sum\_Metric$ of 12 groups of REFs}} 
	\begin{tabular}{lcccc}
		\hline
		\diagbox{\qquad R} {${Sum\_Metric}_{M}^{R}$} {M}
		& FMI             & JC              & PR              & RR              \\ \hline
		\textbf{Group1}  & 173.71          & 153.84          & 174.06          & 131.56           \\
		\textbf{Group2}  & 187.3           & 166.28          & 190.2           & 144.81           \\
		\textbf{Group3}  & 128.32           & 116.7           & 122.89           & 108.87           \\
		\textbf{Group4}  & 90.11           & 82.65           & 89.7           & 78.22           \\
		\textbf{Group5}  & 205.81          & 182.04          & 194.5          & 157.79          \\
		\textbf{Group6}  & 193.78          & 170.71           & 192.12          & 151.15           \\
		\textbf{Group7}  & 164.79          & 146.17           & 164.8          & 125.25           \\
		\textbf{Group8}  & 140.31           & 123.99          & 135.45           & 107.17           \\
		\textbf{Group9}  & 196.54          & 172.78         & 194.8          & 152.4              \\
		\textbf{Group10} & 150.27          & 133.76           & 147.51           & 117.59            \\
		\textbf{Group11} & 206.48          & 181.21          & 205.38          & 171.01          \\
		\textbf{Group12} & \textbf{221.33} & \textbf{196.62} & \textbf{227.15} & \textbf{177.47} \\ \hline
	\end{tabular}
\end{table}

For two REFs, $R_1$ and $R_2$, if $Sum\_Metric_M^{R_1}$  $\textgreater$ $Sum\_Metric_M^{R_2}$, it means that according to the metric $M$, $R_1$ is better than $R_2$ in representing failed test cases. We contrast 12 groups of REFs according to their ${Sum\_Metric}_{M}^{R}$ in Table \ref{tab:contrast}\footnote{In the cell of [$R_1$, $R_2$], {\color[HTML]{2D6AA7} $\checkmark$}, {\color[HTML]{5AC5C6} $\checkmark$}, {\color[HTML]{44972B} $\checkmark$}, {\color[HTML]{F19E37} $\checkmark$}indicate that REF $R_1$ is better than REF $R_2$ in terms of FMI, JC, PR, or RR, respectively.}, as well as list the ${Sum\_Metric}_{M}^{R}$ values in Table \ref{tab:summetric}. It can be seen that Group 12 outperforms the other 11 groups of REFs regardless of being evaluated by FMI, JC, PR, or RR.

\textbf{Now we can draw the conclusion of RQ1:} Group12 is highly competitive across all REFs when representing failed test cases. The list of 12 groups of REFs ranked by their capability to representing failed test cases is as follows:

\begin{center}\footnotesize
	\begin{tcolorbox}[colback=gray!15,%gray background
		colframe=black,% black frame colour
		width=8cm,% Use 8cm total width,
		arc=1mm, auto outer arc,
		boxrule=0.5pt,
		]
		{\textbf{Group12} $>$ Group11 $>$ Group5 $>$ Group9 $>$ Group6 $>$ Group2 \\
			 $>$ Group1 $>$ Group7 $>$ Group10 $>$ Group8 $>$ Group3 $>$ Group4}
	\end{tcolorbox}
\end{center}

\subsection{The impact of NOF contained in PUT on the clustering effectiveness (RQ2)}
\label{subsect4.2}

Similar to definitions depicted in Section \ref{subsect4.1}, we first use $V_{equal}^N$ ($N$ = 2, 3, 4, 5) to denote how many $N$-bug faulty versions' NOF can be accurately estimated by a specific REF, then define and employ the ${Sum\_Metric}_{M}^{N}$, as illustrated in Formula~\ref{equ9}, to observe the clustering effectiveness on these $V_{equal}^N$ versions. 

\begin{equation}
	\label{equ9}
	{Sum\_Metric}_{M}^{N}=\sum_{i}^{V_{equal}^{N}} M_{i}
\end{equation}

The clustering effectiveness is visualized using box-and-whisker plots in terms of upper quartile, lower quartile, median, and mean, where each vertical column's color reflects the value of $V_{equal}^N$ (a darker color indicates a greater value of $V_{equal}^N$). The color is regulated by adjusting the opacity using the procedures below\footnote{\synote{This color setting scheme is also applicable to Section \ref{subsect4.3}.}}

\begin{itemize}
	\item \textbf{Step-1}: Set the color of each vertical column to $black$ (RGB: 0, 0, 0).
	
	\item \textbf{Step-2}: Count the values of $V_{equal}^N$ in $N$-bug scenarios ($N$ = 2, 3, 4, 5), and set the maximum value to $MAX$, as defined in Formula \ref{equ15}.
	
	\begin{equation}
		\label{equ15}
		M A X=\max \left\{V_{equal}^{N}\right\} \quad(N = 2, 3, 4, 5)
	\end{equation}
	
	\item \textbf{Step 3}: Calculate the opacity $Opacity_N$ of each vertical column, as defined in Formula \ref{equ16}.
	
	\begin{equation}
		\label{equ16}
		Opacity_ N = \frac{V_{equal}^{N}}{M A X} \quad(N = 2, 3, 4, 5)
	\end{equation}
	
\end{itemize}

\begin{figure}  
	\centering  
	\includegraphics[height=6cm,width=8cm]{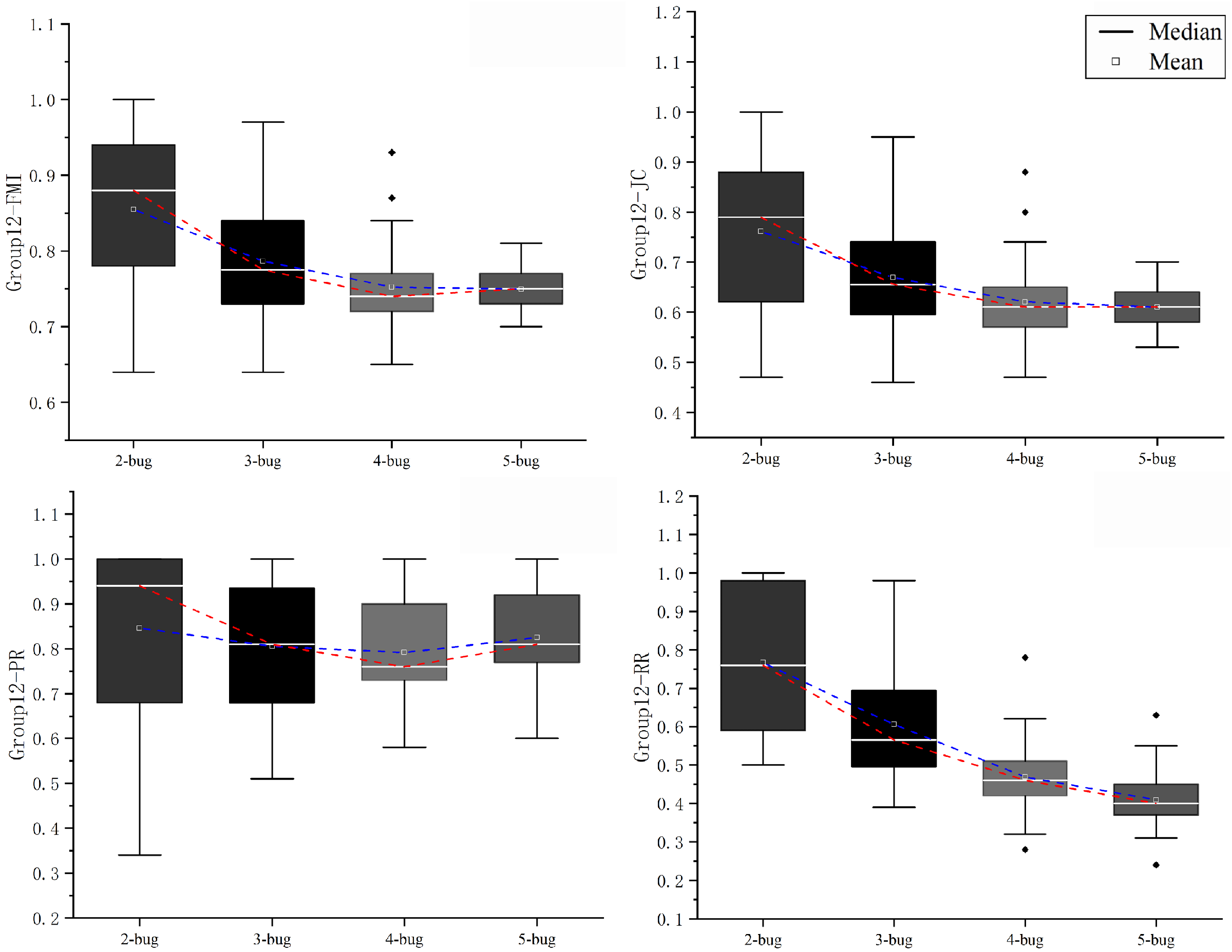}  
	\caption{\synote{The contrast of clustering effectiveness among 2-bug, 3-bug, 4-bug, and 5-bug scenarios}}  
	\label{fig:RQ2}  
\end{figure}

The clustering effectiveness in 2-bug, 3-bug, 4-bug, and 5-bug scenarios is shown in \synote{Figure \ref{fig:RQ2}} \footnote{Due to space limitations, we only display the clustering results of Group12, despite the fact that the clustering results of the other 11 groups of REFs all confirm the conclusions in Section \ref{subsect4.2}. Please refer to \href{https://github.com/yisongy/failureClustering}{the supplementary material} for a complete list of conclusions.}. From this, \textbf{we can draw the conclusions of RQ2:}

\hangafter=1
\setlength{\hangindent}{2.7em}
1) As NOF increases, the similarities (FMI, JC) between generated clusters and oracle clusters decrease.

\hangafter=1
\setlength{\hangindent}{2.7em}
2) As NOF increases, the Recall Rate (RR) falls, while the Precision Rate (PR) changes little.

\hangafter=1
\setlength{\hangindent}{2.7em}
3) As NOF increases, the dispersion of FMI, JC, and RR narrows.

\hangafter=1
\setlength{\hangindent}{2.7em}
4) Based on the ranking lists produced by Group12, a greater value of $V_{equal}^N$ tends to be obtained if $N$ equals 3 (Group3, Group5, Group8, and Group10 also support this conclusion).

The list of NOFs ranked by the clustering effectiveness under them is as follows:

\begin{center}\footnotesize
	\begin{tcolorbox}[colback=gray!15,%gray background
		colframe=black,% black frame colour
		width=4.7cm,% Use 8cm total width,
		arc=1mm, auto outer arc,
		boxrule=0.5pt,
		]
		{\textbf{2-bug} $>$ 3-bug $>$ 4-bug $>$ 5-bug}
	\end{tcolorbox}
\end{center}

\subsection{The impact of FT contained in PUT on the clustering effectiveness (RQ3)}
\label{subsect4.3}

Similar to definitions depicted in Section \ref{subsect4.1}, we first use $V_{equal}^T$ ($T$ takes A, P, H) to denote how many Type$T$ faulty versions' NOF can be accurately estimated by a specific REF, then define and employ the ${Sum\_Metric}_{M}^{T}$, as illustrated in Formula~\ref{equ10}, to observe the clustering effectiveness on these $V_{equal}^T$  versions.

\begin{equation}
	\label{equ10}
	{Sum\_Metric}_{M}^{T}=\sum_{i}^{V_{equal}^{T}} M_{i}
\end{equation}

\begin{figure}  
	\centering  
	\includegraphics[height=6cm,width=8cm]{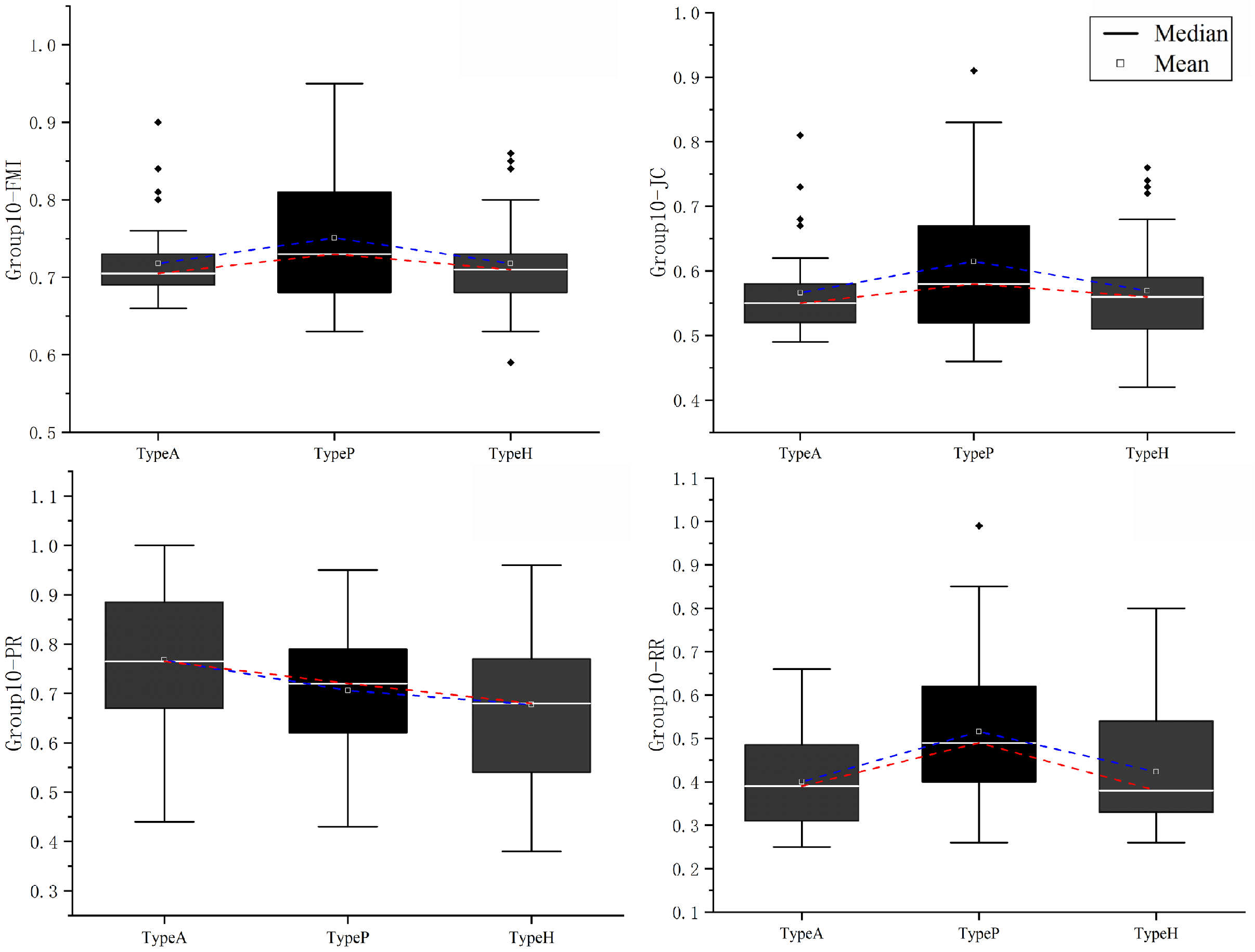}  
	\caption{\synote{The contrast of clustering effectiveness among TypeA, TypeP, and TypeH scenarios}}  
	\label{fig:RQ3}  
\end{figure}

The clustering effectiveness in TypeA, TypeP, and TypeH scenarios is shown in \synote{Figure \ref{fig:RQ3}} \footnote{Due to space limitations, we only display the clustering results of Group10, despite the fact that the clustering results of many of the other 11 groups of REFs confirm the conclusions in Section \ref{subsect4.3}. Please refer to \href{https://github.com/yisongy/failureClustering}{the supplementary material} for a complete list of conclusions.}. From this, \textbf{we can draw the conclusions of RQ3:}

\hangafter=1
\setlength{\hangindent}{2.7em}
1)	Compared with TypeA and TypeH, better clustering effectiveness is easier to obtain in the TypeP scenario concerning FMI, JC, and RR. No significant differences in terms of PR among the three scenarios are observed.

\hangafter=1
\setlength{\hangindent}{2.7em}
2)	The values of $V_{equal}^T$ and $T$ have no evident relations.

The list of FTs ranked by the clustering effectiveness under them is as follows:

\begin{center}\footnotesize
	\begin{tcolorbox}[colback=gray!15,%gray background
		colframe=black,% black frame colour
		width=4cm,% Use 8cm total width,
		arc=1mm, auto outer arc,
		boxrule=0.5pt,
		]
		{\textbf{TypeP} $>$ TypeA $\approx$ TypeH}
	\end{tcolorbox}
\end{center}

\subsection{The impact of NSP1F on the clustering effectiveness (RQ4)}
\label{subsect4.4}

Unlike the first three RQs in which we pair one failed test case with all (i.e., 100\%) successful test cases, we randomly sample $X$\% ($X$ = 80, 60, 40, 20) of successful test cases to pair with one failed test case in this RQ. Similar to definitions depicted in Section \ref{subsect4.1}, we first use $V_{equal}^X$ to denote how many faulty versions' NOF can be accurately estimated by a specific REF when the proportion of successful test cases is set to $X$\%, then define and employ the ${Sum\_Metric}_{M}^{X}$, as illustrated in Formula \ref{equ11}, to observe the clustering effectiveness on these $V_{equal}^X$ versions.

\begin{equation}
	\label{equ11}
	{Sum\_Metric}_{M}^{X}=\sum_{i}^{V_{equal}^{X}} M_{i}
\end{equation}

\begin{table}[]\footnotesize
	\centering
	\setlength{\belowcaptionskip}{3pt}
	\caption{\label{tab:RQ4} \synote{The contrast of clustering effectiveness among various NSP1Fs}} 
	\begin{tabular}{clccccc}
		\hline
		\multicolumn{2}{c}{\diagbox{\qquad M} {the values of $M$} {X}}
		& 100\%             & 80\%              & 60\%              & 40\%   & 20\%              \\ \hline
		\multirow{2}{*}{FMI}  & mean & 0.82           & 0.82        & 0.82     & 0.81   & 0.81          \\
	    & median & 0.79 & 0.79 & 0.79 & 0.79  & 0.79 \\
		\multirow{2}{*}{JC} &mean &  0.72         & 0.71        & 0.71     & 0.71    &   0.70          \\
		& median & 0.67 & 0.66 & 0.67 & 0.66 & 0.66 \\
		\multirow{2}{*}{PR} &mean & 0.82          & 0.81        &   0.81       & 0.81   &  0.80            \\
		& median & 0.85 & 0.82 & 0.80 & 0.81 & 0.79\\
		\multirow{2}{*}{RR} &mean &  0.68         & 0.67         & 0.66         & 0.67    &  0.65           \\ 
		& median & 0.64 & 0.61 & 0.59 & 0.60& 0.57 \\ 
		\multicolumn{2}{c}{\bm{$\ \ \ \ \ \ \ V_{equal}^X$}} & \textbf{251} &  \textbf{255} & \textbf{261} & \textbf{261} & \textbf{263} \\
		\hline
	\end{tabular}
\end{table}

The clustering effectiveness when $X$ is set to 100, 80, 60, 40, 20 is shown in \synote{Table \ref{tab:RQ4}} \footnote{Due to space limitations, we only display the clustering results of Group11, despite the fact that the clustering results of the other 11 groups of REFs all confirm the conclusions in Section \ref{subsect4.4}. Please refer to \href{https://github.com/yisongy/failureClustering}{the supplementary material} for a complete list of conclusions.}. \synote{For example, ``FMI-mean-80\%: 0.82'' implies that when pairing one failed test case with 80\% of successful test cases, the mean of the values of $FMI$ on 255 ``$k == r$'' faulty versions is 0.82.} From this, \textbf{we can draw the conclusions of RQ4:}

\hangafter=1
\setlength{\hangindent}{2.7em}
1) Lowering NSP1F (to as low as 20\%) has no evident effect on clustering effectiveness.

\hangafter=1
\setlength{\hangindent}{2.7em}
2) The effect of $X$ on the value of $V_{equal}^X$ is neither evident nor decisive.

The list of NSP1Fs ranked by clustering effectiveness under them is as follows:

\begin{center}\footnotesize
	\begin{tcolorbox}[colback=gray!15,%gray background
		colframe=black,% black frame colour
		width=5cm,% Use 8cm total width,
		arc=1mm, auto outer arc,
		boxrule=0.5pt,
		]
		{100\% $\approx$ 80\% $\approx$ 60\% $\approx$ 40\% $\approx$ \textbf{20\%}}
	\end{tcolorbox}
\end{center}

This conclusion indicates that 100\% clustering effectiveness can be achieved with only 20\% of successful test cases. When developers use SRR to clustering failed test cases in parallel debugging, they can feel free to cut the scale of successful test cases for lower debugging costs without worrying about the loss of effectiveness.

\section{DISCUSSION}
\label{sect5}

Some interesting topics related to our empirical study are further discussed in this section.

\subsection{An in-depth analysis of clustering failed test cases}
\label{subsect5.1}

Given a TS and a PUT, the numbers of failed test cases and successful test cases will be immediately determined. If multiple faults are contained in the PUT, all existing failed test cases might be caused by different faults, that is, each failed test case will be linked to its root cause(s). The more the faults, the lower proportion of failed test cases caused by each fault to all failed test cases\footnote{We discuss this problem under the condition of the number of failed test cases has been determined.}. However, the intuition of designing risk evaluation formulas in SBFL is to assign higher suspiciousness to statements that are covered by more failed test cases~\cite{tang2017accuracy[67], pang2015debugging[68]}, which would be disturbed by the presence of multiple faults, and the degree of disturbance magnifies as the number of faults increases. Zheng et al. presented a similar opinion in~\cite{zheng2018localizing[24]}, they claimed when there is only one faulty statement, it is more likely to be covered by more failing executions, whereas the failing executions are diluted by multiple faults so less accurate results are obtained.

To tackle this challenge, it is natural to categorize failed test cases according to their root cause(s), in other words, build linkages between failed test cases and faults. As a classic technique for unsupervised data grouping, clustering is typically employed to accomplish this failure indexing process, with the goal of fault isolation. 

\begin{figure}  
	\centering  
	\includegraphics[height=5.3cm,width=7.8cm]{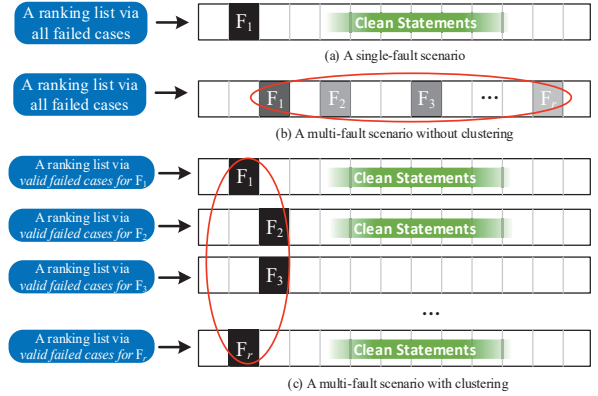}  
	\caption{Fault localization effectiveness with and without clustering in a multi-fault scenario.}  
	\label{fig:singleVSmulti}  
\end{figure}

We use Figure \ref{fig:singleVSmulti} to simulate the effectiveness of fault isolation. In a single-fault scenario, the proportion of failed test cases caused by the unique fault $F_1$ (denoted as valid failed test cases for $F_1$) to all failed test cases is 100\%, that's to say, all failed test cases fed into a risk evaluation formula for $F_1$, thus SBFL techniques are easier to push the statement that contains $F_1$ towards the top of the ranking list, as shown in Figure \ref{fig:singleVSmulti}(a). In a multi-fault scenario, assume there are $r$ bugs, $F_i$ ($i$ = 1, 2, ..., $r$) in a PUT, $n$ failed test cases in a TS, and the number of failed test cases caused by $F_i$ (denoted as valid failed test cases for $F_i$) is $|F_i|$. The proportion of failed test cases linked to $F_i$ to all failed test cases is $|F_i| / n$, which is ordinarily less than 100\%. Furthermore, if all failed test cases are utilized in SBFL without being refined, the process of localizing a single fault, $F_i$, will be interrupted by failed test cases caused by the other faults (denoted as redundant failed test cases for $F_i$). Consequently, SBFL techniques' capability is diminished since linkages between a single fault and its responsible failed test cases have \textbf{been diluted} (simulated by the opacity of faulty statements in Figure \ref{fig:singleVSmulti}(b)), potentially lowering the rankings of statements that contain faults.

After all failed test cases are divided into several disjoint fault-focused clusters, only failed test cases triggered by $F_i$, as well as successful test cases, will be fed into a risk evaluation formula to localize $F_i$. That's to say, when ideal clustering results are delivered, the proportion of valid failed test cases for $F_i$ to all failed test cases \textbf{regains} 100\%, since redundant failed test cases for $F_i$ have been indexed to their own root cause, which enables the position of the statement that contains $F_i$ to be higher in the corresponding ranking list, as shown in Figure~\ref{fig:singleVSmulti}(c).

\subsection{Revisit of $V_{over}^R$ and $V_{under}^R$}
\label{subsect5.2}

When evaluating the capability of REFs to representing failed test cases, we consider only faulty versions that fall into the $Equal$ category, in other words, if the NOF of a faulty version is not accurately estimated based on an REF $R$ (i.e., falls into the $Under$ or the $Over$ category), this faulty version will be discarded, and thus will not be dedicated to $R$'s capability to clustering failed test cases. It is obvious that the larger the value of $V_{equal}^R$ (that is, the lower the values of $V_{over}^R$ and  $V_{under}^R$), the greater the possibility that $R$ will be highly competitive.

Nonetheless, the same values of $V_{over}^R$ and $V_{under}^R$ should not be treated equally since they can reflect different deviations from the NOF. For example, assume that the NOFs of ten 5-bug faulty versions are being estimated based on the ranking lists produced by two REFs, $R_1$ and $R_2$, respectively, we can immediately get $r_i$ ($i$ = 1, 2, …, 10) are all equal to 5. If the estimate results generated by $R_1$ are $k_i^{R_1}$ ($i$ = 1, 2, …, 10), which are 9, 9, 8, 9, 5, 5, 1, 1, 2, 2, respectively, and the estimate results generated by $R_2$ are $k_i^{R_2}$ ($i$ = 1, 2, …, 10), which are 6, 6, 7, 6, 5, 5, 3, 3, 4, 4, respectively. According to the preceding definitions in Section \ref{subsect4.1}, the values of $V_{over}^{R_1}$ and $V_{over}^{R_2}$ are equal to 4, the values of $V_{equal}^{R_1}$ and $V_{equal}^{R_2}$ are equal to 2, and the values of $V_{under}^{R_1}$ and $V_{under}^{R_2}$ are equal to 4. Although both $R_1$ and $R_2$ estimate the NOF on eight faulty versions inaccurately, it is visible that $R_2$ delivers a closer result, implying $R_2$ has a stronger capability to representing failed test cases to some extent. We define two metrics, $Deviation_{over}^R$ in Formula \ref{equ12} and $Deviation_{under}^R$ in Formula~\ref{equ13}, to quantify this type of difference among all REFs. 

\begin{equation}
	\label{equ12}
	Deviation_{over}^{R}=\frac{1}{V_{over}^{R}} \times \sum_{i}^{V_{over}^{R}}\left(k_{i}-r_{i}\right)
\end{equation}

\begin{equation}
	\label{equ13}
	Deviation_{under}^{R}=\frac{1}{V_{under}^{R}} \times \sum_{i}^{V_{under}^{R}}\left(r_{i}-k_{i}\right)
\end{equation}

Where $k_i$ is the estimated number of clusters on the $i^{\rm{th}}$ faulty version, and $r_i$ represents the NOF contained in the $i^{\rm{th}}$ faulty version.

Using Formula \ref{equ12} and Formula \ref{equ13} to contrast $R_1$ and $R_2$ in the aforementioned example, we can get $Deviation_{over}^{R_1}$ = 3.75, $Deviation_{under}^{R_1}$ = 3.5; $Deviation_{over}^{R_2}$ = 1.25, $Deviation_{under}^{R_2}$ = 1.5. Hence, the difference between $R_1$ and $R_2$ hidden behind the $k \neq r$ faulty versions is captured and quantified.

We revisit the values of $V_{over}^R$ and $V_{under}^R$ for 12 groups of REFs presented in Figure \ref{fig:equalversion}, as shown in Table \ref{tab:revisit}.

\begin{table}[]\footnotesize
	\centering
	\setlength{\belowcaptionskip}{3pt}
	\caption{\label{tab:revisit} \synote{The values of $Deviation$ of 12 groups of REF}}
	\begin{tabular}{cccc}
		\hline
		\diagbox{\tiny{R}}{\tiny{Value}}{\tiny{Metrics}} & $Deviation_{over}^{R}$ & $Deviation_{under}^{R}$ & $mean$ \\ \hline
		\textbf{Group1}               & 3.32 & \textbf{1.57} & 2.95 \\
		\textbf{Group2}               & 2.89 & 1.70  & 2.53 \\
		\textbf{Group3}               & 4.81 & 1.81 & 3.72 \\
		\textbf{Group4}               & 1.95 & 2.39 & 2.36 \\
		\textbf{Group5}               & 2.02 & 1.73 & \textbf{1.87} \\
		\textbf{Group6}               & 2.37  & 1.85 & 2.13 \\
		\textbf{Group7}               & 3.46 & 1.66 & 3.09 \\
		\textbf{Group8}               & 4.21 & 2.15 & 3.15 \\
		\textbf{Group9}               & 2.35 & 1.87 & 2.13 \\
		\textbf{Group10}              & 3.80 & 2.04 & 3.20 \\
		\textbf{Group11}              & \textbf{1.26} & 2.05 & 1.94 \\
		\textbf{Group12}              & 2.07  & 1.80 & 1.95 \\ \hline
	\end{tabular}
\end{table}

It can be seen that the value of $Deviation_{over}^{R}$ of Group11 is \synote{1.26}, indicating when the estimated number of clusters exceeds the NOF, Group11 has the lowest degree of $over$-$representing$. The value of $Deviation_{under}^{R}$ of Group1 is \synote{1.57}, indicating when the estimated number of clusters is fewer than the NOF, Group1 has the lowest degree of $under$-$representation$. The $mean$ of \synote{Group5} is \synote{1.87}, indicating when the estimated number of clusters is not equal to the NOF, \synote{Group5} has the lowest deviation.

\minorrevised{Notice that such analyses are non-trivial for parallel debugging. In real multi-fault localization scenarios, it is expected that the predicted number of faults $k$ is identical to the number of faults $r$. If such ideal situations cannot be attained, the smaller the deviation, the lower the time and labor cost. Specifically, one cannot judge whether the prediction result is correct since the value of $r$ is unknown in practice. Thus, $k$ fault-focused clusters will be directly input to the following localization stage. If $k$ exceeds $r$, $k$ developers will be employed to locate $r$ faults, resulting in waste of human labor ($k$~-~$r$ developers are redundant). On the contrary, if $k$ is less than $r$, more than one ($\lceil r / k \rceil$) iteration of debugging is needed, resulting in waste of time.}

\subsection{A heuristic perspective to contrast REFs}
\label{subsect5.3}

We further discuss the relation between the virtual mapping problem and the evaluation of clustering effectiveness. Assume that REF $R$ is utilized to represent failed test cases in a faulty version. If the estimated number of clusters $k$ is equal to the NOF $r$, there will be $A_k^r$ permutations between generated clusters and oracle clusters. The four metrics, FMI, JC, PR, and RR, will appear different values on different permutations. If the highest values of the four metrics all appear on the same permutation, it means that the four metrics can easily achieve a consensus, which indicates that the ranking lists produced by $R$ represent failed test cases distinguishably. On the contrary, if the highest values of the four metrics are dispersed onto different permutations, divergences among these four metrics are revealed, which just demonstrates that the ranking lists produced by $R$ are too analogous to be divided.

We regard the evaluation of four metrics for all permutations as a voting process, in which each metric votes for the permutation with its highest value. For example, a permutation will get four votes if the highest values of all four metrics occur on it. Obviously, in the aforementioned $r$-bug faulty version, $A_k^r$ permutations will each be assigned a value of votes. This $r$-bug faulty version's votes will be referred to as the highest value of votes among $A_k^r$ permutations.

We design the $Sum\_Vote^{R}$ metric to count the votes of faulty versions that satisfy the ``$k$ == $r$" criteria for each REF $R$ in Figure \ref{fig:equalversion}, as shown in Formula \ref{equ14}. We believe that the $Sum\_Vote^{R}$ metric reflects the capability of the risk evaluation formula $R$ to representing failed test cases from a heuristic perspective.

\begin{equation}
	\label{equ14}
	{Sum\_Vote}^{R}=\sum_{i}^{V_{equal}^{R}} vote_{i}
\end{equation}

Where $vote_i$ is the value of votes of the $i^{\rm{th}}$ faulty version.

\begin{figure}  
	\centering  
	\includegraphics[height=6.2cm,width=8.4cm]{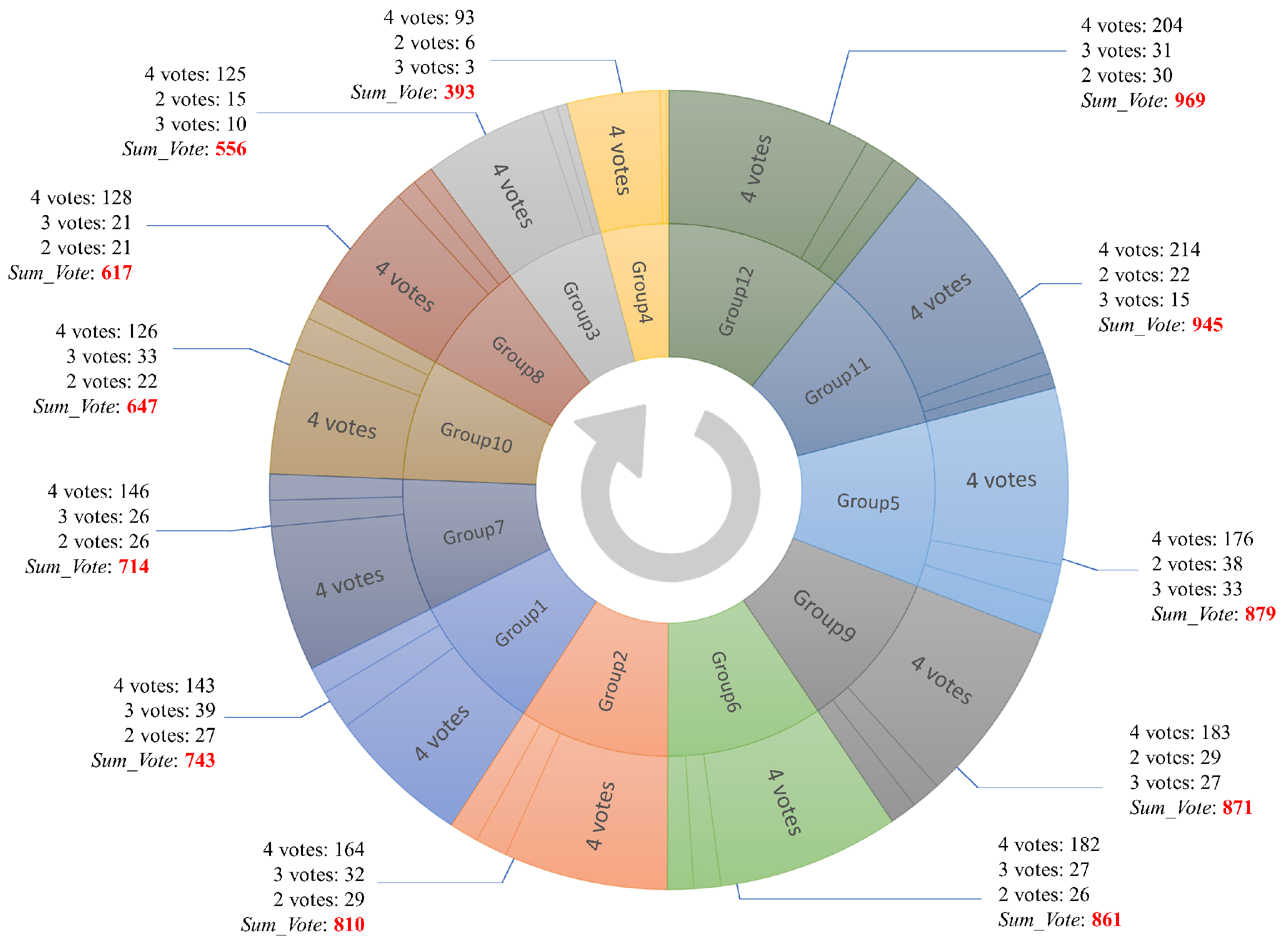}  
	\caption{\synote{The values of $Sum\_Vote^{R}$ of 12 groups of REFs}} 
	\label{fig:heuristic}  
\end{figure}

The values of $Sum\_Vote^{R}$ of 12 groups of REFs are given in Figure \ref{fig:heuristic}. For instance, on \synote{265} ``$k$ == $r$" faulty versions of Group12,  \synote{204}, 31, and 30 of them get 4, 3, and 2 votes, respectively, we can immediately obtain $Sum\_Vote^{Group12}$ = \synote{969} according to Formula \ref{equ14}. The direction of the circular arrow in Figure \ref{fig:heuristic} indicates the ranking of $Sum\_Vote^{R}$ values of 12 groups of REFs: Group12 $>$ Group11 $>$ Group5 $>$ Group9 $>$ Group6 $>$ Group2 $>$ Group1 $>$ Group7 $>$ Group10 $>$ Group8 $>$ Group3 $>$ Group4, double-confirming the conclusion of RQ1.

\subsection{Why is it easier to obtain better clustering effectiveness in TypeP faulty versions?}
\label{subsect5.4}

The conclusions in Section \ref{subsect4.3} reveal that when a program has only predicate faults, the overall clustering effectiveness is higher than when it has only assignment faults and both two types of faults coexist. \synote{Take Group10 as an example (Figure~\ref{fig:RQ3}), the number of ``$k == r$'' faulty versions of TypeP is 25.0\% and 28.6\% greater than that of TypeA and TypeH, respectively, according to their opacity. TypeP scenarios also have better clustering effectiveness (the mean and median of FMI, JC, and RR) than the other two fault types.}

In SRR-based failure clustering, a ranking list, which is produced by a risk evaluation formula, serves as a proxy for a failed test case. The basis of generating a ranking list is spectrum information, while the latter originates from coverage. In other words, SRR-based failure clustering heavily depends on the failed test cases' execution paths on the PUT. For failed test cases caused by different faults, the more distinctive execution paths they have, the more distinguishable ranking lists an REF can generate, and the easier they are to be indexed. \synote{A TypeP faulty version has only predicate faults, which involve reversing the $if$-$else$ predicate, deleting the $else$ statement, or modifying the decision condition, etc., according to the definition in Section \ref{subsubsect:sir}. All of the three classes could cause unwanted code to be executed, resulting in a different trace.} Thus, failed test cases \synote{in TypeP faulty versions} are more likely to appear diverse coverage, which will be beneficial to isolate \synote{these predicate faults}. However, this assistance, on the one hand, does not exist when a program contains only assignment fault, on the other hand, is diminished when the two types of faults coexist.

\subsection{The function of successful test cases in SRR}
\label{subsect5.5}

In Section \ref{subsubsect3.2.4}, we assume that the function of successful test cases in SRR-based failure clustering is to assist risk evaluation formulas in generating ranking lists (some REFs will lose their definition without being fed into successful test cases), that's to say, they serve as complements in failure indexing. The conclusions in Section \ref{subsect4.4} reveal that lowering NSP1F (to as low as 20\%) indeed has no evident effect on clustering effectiveness. As a result, while performing SRR-based failure clustering, developers can reduce debugging costs by pairing only a portion of the successful test cases with one failed test case since too many successful test cases will not help represent failed test cases.

Even though failed test cases have gotten a lot of attention in testing and debugging, successful test cases can also play a vital role. For example, metamorphic testing enables successful test cases to expose failures via metamorphic relations~\cite{chen2020metamorphic[49], xie2013metamorphic[50]}. We only illustrate the redundancy of successful test cases in SRR-based failure clustering, without denying their significance in localization, testing, or the other software quality assurance activities.

\section{THREATS TO VALIDITY}
\label{sect6}

Similar to previous empirical studies on parallel debugging, a hard-clustering strategy is used in this paper to divide failed test cases, that is, a failed test case can only be categorized into one cluster. However, in real-world debugging processes, the relations between faults and failures are quite complex since several faults might trigger the same failure (i.e., one failed test case links to multiple faults). Therefore, the clustering effectiveness will be reduced since the inherent conflict between the property of hard-clustering techniques and the one-to-many or many-to-many linkages. Nonetheless, the reliability of our conclusions is not affected by this threat since we contrast different variables based on the same clustering technique.

In addition, to build the virtual linkages between generated clusters and oracle clusters, we filter out faulty versions with the estimated number of clusters not equal to the NOF. Although this strategy guarantees the availability of clustering results, it also causes various variables in each RQ to be contrasted based on different numbers of faulty versions. This threat seems to introduce additional uncertainties for the experiments, however, we believe that 1) how many faulty versions are selected by various variables in each RQ (i.e., 12 groups of REFs in RQ1, 2-bug, 3-bug, 4-bug, and 5-bug scenarios in RQ2, TypeA, TypeP, and TypeH scenarios in RQ3, 100\%, 80\%, 60\%, 40\%, and 20\% of successful test cases in RQ4) reflect these variables' capability to representing failed test cases, and 2) the distinction in diverse benchmarks avoids the bias caused by a standard dataset, which makes the conclusions more universal.

Although we collected four datasets with varied scales and functions, they are all written in C. Besides, when utilizing the mutation-based strategy to inject faults into the original program, the number of predefined mutation operators is limited, which lowers the diversity of faulty versions to some extent.

\section{RELATED WORK}
\label{sect7}

Clustering failed test cases into various fault-focused groups that target different faults is not a newborn method. As early as 2003, Podgurski et al. observed that open-source software developers had received a large number of bug reports from end-users every day, but many of these bug reports are actually caused by the same fault although they have distinct trigger paths and different anomalous behaviors. To that end, they suggested grouping together failures with the same root cause based on supervised and unsupervised pattern classification, which avoids potentially unwanted and redundant debugging labor~\cite{podgurski2003automated[15]}.  Considering the suggestions of Podgurski et al.~\cite{podgurski2003automated[15]}, Jones et al. proposed two parallel debugging techniques in~\cite{jones2007debugging[2]}. Specifically, they first divided failed test cases into several disjoint clusters based on similarities, and then separately combined these clusters with all successful test cases to generate specialized test suites that are expected to target different faults. These fault-focused TSs are finally assigned to several developers for localizing multiple faults in parallel. DiGiuseppe and Jones then conducted an empirical study to confirm the necessity of clustering failed test cases and to explore the influence of the presence of multiple faults on fault localization. They pointed out clustering failed test cases is necessary and beneficial despite the fact that this process may incur additional computational costs, since their findings demonstrated that multi-fault indeed had a negligible effect on the effectiveness of fault localization~\cite{digiuseppe2011influence[4]}.

Högerle et al. first quoted an important opinion concluded by Jones et al. in~\cite{jones2007debugging[2]}, that is, parallelization can speed up debugging significantly, even if the derived parallel tasks are conducted sequentially, and then pointed out that the method of dividing failed test cases should be carefully chosen because it will have a significant impact on the division effectiveness through large-scale experiments~\cite{hogerle2014more[16]}. The effectiveness of parallel debugging will be directly determined by the outcomes of the clustering process. Zakari and Lee investigated commonly-used parallel debugging techniques and found that most research 1) employed CVR as failure proximity to represent failed test cases, and 2) used Euclidean, Jaccard, or Hamming distance to measure the similarities between failed test cases. They first coined the term $problematic\ approach$ to describe debugging approaches that adopted the above techniques, and then conducted an empirical study on the effectiveness of several problematic approaches adopting the K-means clustering algorithm. Their results showed that clustering built upon CVR and Euclidean distance reduced the effectiveness of multi-fault localization~\cite{zakari2019parallel[17]}.

Liu et al. conducted systematic research on failure proximity in~\cite{liu2008systematic[6]} and~\cite{liu2006failure[76]}, in which they summarized or proposed six representative failure proximities, i.e., Failure-based, Stack Trace-based, Code Coverage-based, Predicate Evaluation-based, Dynamic Slicing-based, and Statistical Debugging-based. The CVR utilized in most studies is similar to the above-mentioned Trace-proximity, which has been proven to be less effective in clustering failed test cases. To tackle this limitation, Gao and Wong employed SRR, which is similar to Rank-proximity in~\cite{liu2008systematic[6]}, to represent failed test cases. Specifically, 1) they paired each failed test case with all successful test cases and input them into an REF, Crosstab~\cite{wong2011towards[18]}, to generate a ranking list that represents the corresponding failed test case. 2) They stated that the clustering algorithm's performance highly depends on the distance metric, thus revised the original Kendall tau distance based on the premise that discordant pairs of more suspicious statements contribute more to the distance between two ranking lists. 3) To tackle the long-standing problem of estimating the number of clusters, as well as relieving the uncertainty introduced by randomly generating initial centroids, an approach of selecting initial medoids while predicting the number of clusters was presented inspired by prior studies~\cite{yager1994approximate[19], chiu1994fuzzy[20]}. 4) They claimed that their initial medoids selection approach reduced the high computational costs to a large extent compared with the original K-medoids clustering algorithm, due to the latter examines all possible combinations of data points as initial medoids. Gao and Wong integrated the above four innovations and developed a novel technique for localizing multiple faults in parallel~\cite{gao2017mseer[14]}.

In addition, some researchers have developed a series of novel parallel debugging strategies by integrating techniques from other domains into fault localization. For example, Zakari et al. proposed a fault localization technique that is suited for both single-fault and multi-fault scenarios based on the complex network theory (FLCN), where developers can localize multiple faults at the same time in a single diagnosis ranking list~\cite{zakari2018simultaneous[21]}. In another study, they adopted the divisive network community algorithm to cluster failed test cases, as well as employed a weighting and selecting mechanism to prioritize generated fault-focused communities~\cite{zakari2019community[22]}. Based on one-fault-at-a-time via OPTICS (Ordering Points To Identify the Clustering Structure) clustering, Wu et al. proposed to 1) divide failed test cases in each iteration and calculate the density of each cluster, 2) combine the failed test cases in the cluster with the highest density value with all successful test cases to form a new test suite, and 3) localize a single fault based on the ranking list produced by the new test suite, iterating these steps until all bugs are fixed. Based on their findings, they further concluded that using the clustering algorithm with the highest accuracy can achieve the best performance of multi-fault localization~\cite{wu2020fatoc[23]}. Inspired by the multiple-fault-at-a-time strategy, Zheng et al. converted fault localization tasks into search problems and proposed a fast software multi-fault localization framework using genetic algorithms~\cite{zheng2018localizing[24]}. Pei et al. introduced the dynamic random testing (DRT) strategy and proposed distance-based DRT, which vectorized test cases and divided them into disjoint subdomains using distance information from inputs and a specific clustering algorithm~\cite{Pei2021[57]}.

There are also some researchers who carried out empirical comparisons of different techniques in the field of multi-fault localization. For instance, Gao et al. contrasted the effectiveness of 22 machine learning algorithms typically used in multi-fault localization and found that random forests, BP neural networks, and logit boost machine learning models based on ensemble learning performed well~\cite{gao2018research[25]}. Huang et al. first created 12 types of setup by combining 6 REFs and 2 widely-used clustering algorithms, and then conducted empirical research in multi-fault scenarios using CVR. Their experimental results showed that Wong1 paired with K-means outperformed the other combinations~\cite{huang2013empirical[26]}. Zakari et al. conducted a systematic literature review on classic parallel debugging techniques~\cite{zakari2020multiple[27]}. They investigated off-the-shelf studies and categorized them into three prominent types of strategy, one-fault-at-a-time debugging, parallel debugging, and multiple-fault-at-a-time debugging. Among them, they pointed out parallel debugging alleviated fault interferences through clustering failed test cases. However, many studies such as~\cite{jones2007debugging[2]} and~\cite{huang2013empirical[26]} claimed these existing strategies were insufficient for isolating faults as well as listed some challenges related to clustering effectiveness in parallel debugging, including the method of representing failed test cases, the initial set of fault-focused clusters, the clustering algorithm, and the distance metric.

\section{CONCLUSION AND FUTURE WORK}
\label{sect8}

We extract and analyze four essential factors, i.e., the risk evaluation formula that produces ranking lists, the number of faults in a program, the fault types, and the number of successful test cases paired with one individual  failed test case, to investigate how these variables affect clustering effectiveness. Four research questions are presented in this paper, the corresponding controlled experiments show that: 1) GP19 is highly competitive across all REFs, thus we recommend that researchers or developers who adopt SRR for parallel debugging use GP19 to represent failed test cases; 2) clustering effectiveness decreases as NOF increases, indicating that a greater number of faults reduces the effectiveness not only in fault localization but also in fault isolation; 3) higher clustering effectiveness is easier to achieve when a program contains only predicate faults, which points out the challenge of isolating assignment faults; and 4) clustering effectiveness remains when NSP1F is reduced to 20\%, future researchers and developers are suggested to cut the scale of successful test cases while using SRR for a lower debugging expense.

In the future, we plan to further explore the internal mechanisms of risk evaluation formulas to representing failed test cases, followed by proposing a novel REF for the representation of failed test cases. We also consider investigating the four factors that may influence clustering effectiveness with larger datasets and broader experiment setups, as well as introducing new evaluation metrics.

\section*{Acknowledgment}
This work was partially supported by \iffalse the National Key R\&D Program of China under the grant number 2020AAA0107803, and \fi the National Natural Science Foundation of China under the grant numbers 61972289 and 61832009. And the numerical calculations in this work have been partially done on the supercomputing system in the Supercomputing Center of Wuhan University.

\bibliography{ref}
\end{CJK*}
\end{document}